\documentclass[journal=jpclcd,manuscript=article,layout=onecolumn,traditional]{achemso}
\setkeys{acs}{usetitle=true, doi=true}

\usepackage{chemformula} % Formula subscripts using \ch{}
\usepackage[T1]{fontenc} % Use modern font encodings

\usepackage{amsmath,amssymb}
\usepackage[version=3]{mhchem} % chemical typesetting
\usepackage{graphicx}% Include figure files
\graphicspath{{./figures/}{JPCB-followup/}}

\usepackage{bm} % bold math

\usepackage{hyperref}
\hypersetup{colorlinks=true, linkcolor=black,urlcolor=black!50!black, citecolor=black!50!black}

\usepackage[capitalize]{cleveref}
\crefrangelabelformat{equation}{(#3#1#4)$-$(#5#2#6)}

\usepackage[separate-uncertainty]{siunitx}

\usepackage{chemarr}

\usepackage{cuted}

\DeclareMathOperator\Det{Det}
\DeclareMathOperator\Tr{Tr}

\usepackage[normalem]{ulem}

\author{Arthur V. Straube}
\affiliation{%Division of Mathematics for Life and Materials Sciences, \\
Zuse Institute Berlin, Takustra{\ss}e 7, 14195 Berlin, Germany}
\email{straube@zib.de}
\author{Stefanie Winkelmann}
\affiliation{%Division of Mathematics for Life and Materials Sciences, \\
Zuse Institute Berlin, Takustra{\ss}e 7, 14195 Berlin, Germany}

\author{Felix H{\"o}fling}
\affiliation{Freie Universit{\"a}t Berlin, Department of Mathematics and Computer Science, \\ Arnimallee 6, 14195 Berlin, Germany}
\alsoaffiliation{Zuse Institute Berlin, Takustra{\ss}e 7, 14195 Berlin, Germany}

\title[]{Accurate Reduced Models for the pH Oscillations in the Urea--Urease Reaction Confined to Giant Lipid Vesicles}

\begin{document}

\begin{tocentry}
\includegraphics[width=1.0\textwidth]{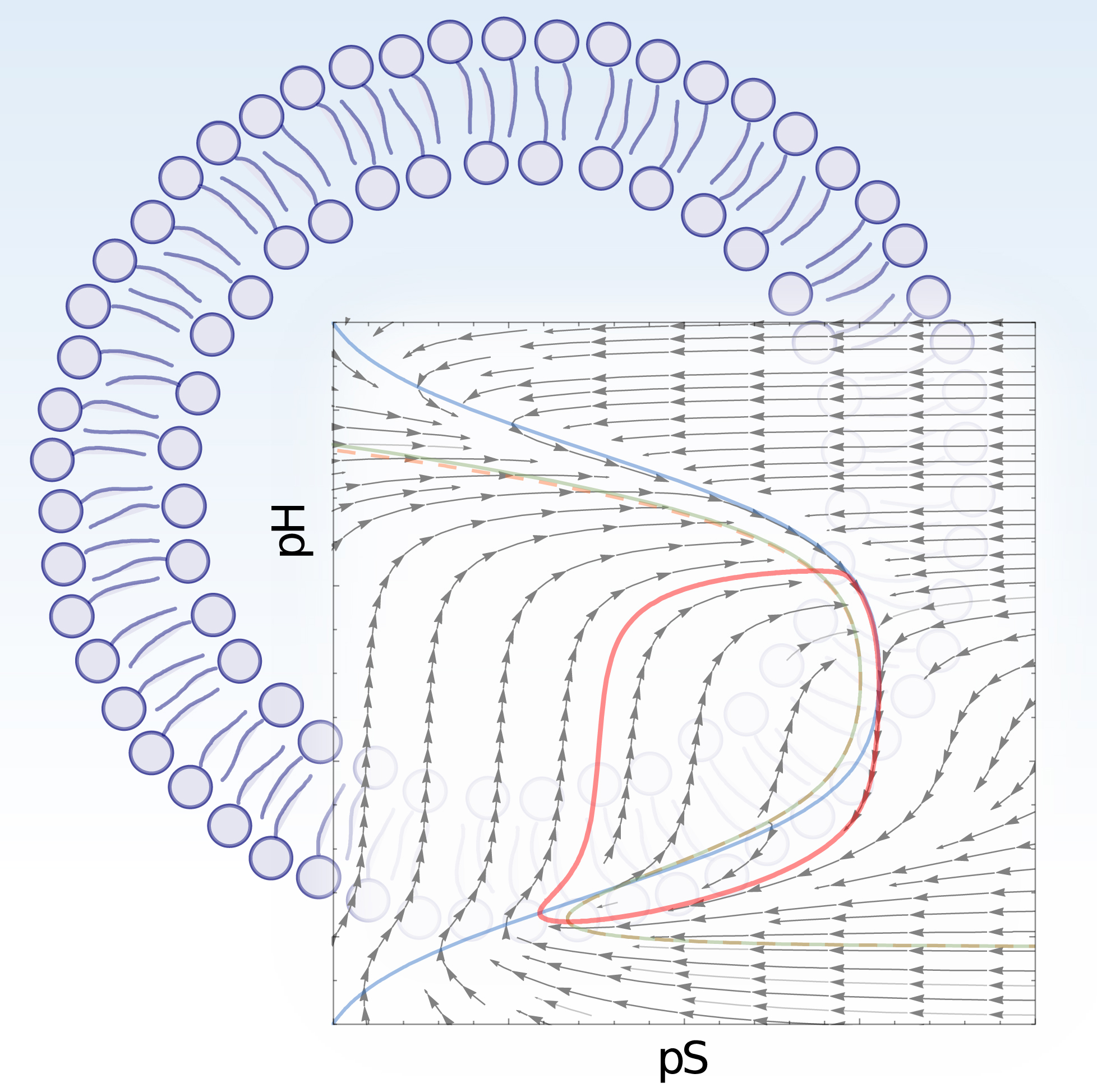}
%The urea–urease reaction confined to a giant lipid vesicle can lead to a perfectly periodic pH oscillations. In our numerical study we show that downscaling to nano-sized vesicles results in significant statistical variation of the oscillation period with a robust mean down to a certain vesicle size, below wich the periodicity of rhythms is gradually destroyed.
\end{tocentry}

\begin{abstract}
This theoretical study concerns a pH oscillator based on the urea--urease reaction- confined to giant lipid vesicles. Under suitable conditions, differential transport of urea and hydrogen ion across the unilamellar vesicle membrane periodically resets the pH clock that switches the system from acid to basic, resulting in self-sustained oscillations. We analyse the structure of the phase flow and of the limit cycle, which controls the dynamics for giant vesicles and dominates the pronouncedly stochastic oscillations in small vesicles of submicrometer size. To this end, we derive reduced models, which are amenable to analytic treatments that are complemented by numerical solutions, and obtain the period and amplitude of the oscillations as well as the parameter domain, where oscillatory behavior persists. We show that the accuracy of these predictions is highly sensitive to the employed reduction scheme. In particular, we suggest an accurate two-variable model and show its equivalence to a three-variable model that admits an interpretation in terms of a chemical reaction network. The faithful modeling of a single pH oscillator appears crucial for rationalizing experiments and understanding communication of vesicles and synchronization of rhythms.
\end{abstract}


\section{Introduction}

Recent years have seen a growing surge of interest in design and development of chemical oscillators for various applications~\cite{Novak:NRMCB2008, Epstein-Vanag-etal:ACR2012, Orban:ACR2015, Cupic:FC2021}. Both in natural intracellular environments and under engineered \textit{in vitro} conditions, the enzyme-assisted reaction kinetics is typically confined to small vesicles, i.e., permeable membrane-based micro- to nano-sized compartments~\cite{Zhang-etal:CR2021}. The concentration of the hydrogen ion, \ce{H+}, or, equivalently, the level of \ce{pH} is an important factor that controls the speed of enzymatic reactions.\cite{Alberty-Massey:BBA1954} Systems in which the hydrogen ion plays the central role and causes self-sustained oscillatory behavior belong to the class of pH oscillators.\cite{Orban:ACR2015} Many examples of \ce{pH} oscillations result from an interplay of chemical reactions that involve positive and negative feedback and occur in closed reactors. In contrast to conventional oscillators, the mechanism of \ce{pH} oscillations discussed here relies on an open reactor.

Motivated by experimental implementations~\cite{Hu-etal:JPCB2010, Muzika-etal:PCCP2019} and direct relevance for applications~\cite{Miele:Proc2016, Miele:LNBE2018, Miele-etal:JPCL2022}, we consider an urea-urease-based pH oscillator confined to a lipid vesicle as an open reactor. Ureases are a group of enzymes for the hydrolysis of urea \cite{Krajewska:JMCBE2009}, which occur widely in the cytoplasm of bacteria, invertebrates, fungi, and plants, but also in soils. The activity of urease is highly sensitive to the pH level and is maximal in a pH-neutral environment \cite{Qin:ABB1994, Fidaleo-Lavecchia:CBEQ2003, Krajewska:PPB2005, Krajewska:JMCBE2009}. This renders the urea--urease reaction a typical \ce{pH} clock that switches the system from acid to basic \cite{Hu-etal:JPCB2010, Bubanja-etal:RKMC2018}.
The clock can be ``reset'' if one allows for the exchange of acid and urea with an external reservoir such that the initial concentrations are recovered, thereby completing the elementary cycle of the oscillator.
One potential realization of such a pH oscillator makes use of differential transport of hydrogen ion and urea across lipid vesicle membranes \cite{Bansagi:JPCB2014}: placing the vesicles in a suitable urea and pH buffer leads to a recovery of the internal concentrations and thus periodic rhythms.

We have recently studied the impact of intrinsic noise on pH oscillations \cite{Straube-etal:JPCL2021}, which becomes progressively important upon decreasing the vesicle size\cite{Winkelmann:2020book}. It was found that the discrete nature of molecules induces a significant statistical variation of the oscillation period in small, nano-sized vesicles. However, the limit cycle of the deterministic rate equations does not only control the dynamics for giant vesicles (of several micrometers in size), but dominates also the strongly stochastic oscillations in small vesicles.
The goal of this work is the analysis of the structure of the phase flow and the limit cycle. To this end, we derive reduced models, amenable to analytic treatments, and show that the quality of predictions is highly sensitive to the choice of the reduction scheme. In particular, we suggest an accurate two-variable model and show its equivalence to a three-variable model that admits an interpretation in terms of a chemical reaction network.

\begin{figure*}[tb]
	\centering
	\includegraphics[width=1.00\textwidth]{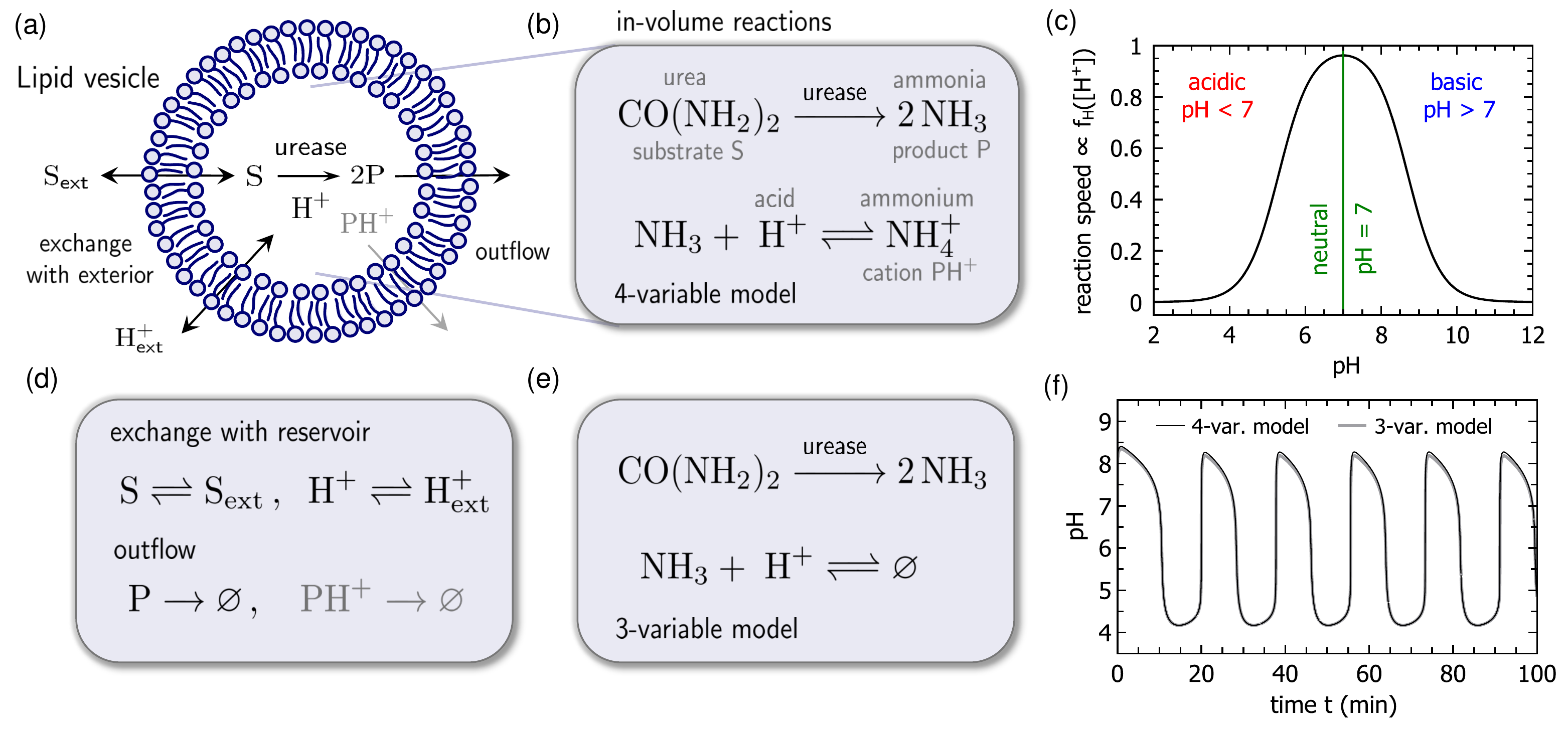}
	\caption{Schematic of the full four-variable and reduced three-variable reaction networks. (a) The enzyme (urease) assists conversion of the substrate \ce{S} (urea) into product \ce{P} (ammonia) in a lipid vesicle affected by varying acidity (hydrogen ion, \ce{H+}). The substrate \ce{S} and acid \ce{H+} exchange with the exterior of the vesicle, whereas the products \ce{P} and \ce{PH+} (ammonium) are subjected to outflow from the vesicle (a,d); the ion form of product \ce{PH+} (reaction shown in gray) belongs only to the full reaction network. 
	In-volume reactions occurring in the vesicle for (b) the full network, \cref{4sm},  and (e) the reduced network, \cref{3sm}.
	(c) The reaction speed of the urease-catalyzed step \eqref{4sm-R1} shows a bell-shaped dependence $f_\textrm{H}([\ce{H}^+])$
	on $\ce{pH}=-\log_{10}([\ce{H+}]/\SI{1}{M})$ with the maximum in a neutral medium ($\ce{pH} = 7$), see \cref{eq:kcat,eq:kcat-MM,eq:fH}.  (f) \ce{pH} oscillations obtained from the numerical solution of the four- and three-variable models, \cref{eq:rre4,eq:rre3}, see also \cref{fig:4-var-model}. 
    }
    \label{fig:sketch}
\end{figure*}

\section{Reaction scheme and four-variable model}

We start with the four-variable model of \ce{pH} oscillations in the urea--urease reaction confined to lipid vesicle applied in our earlier study\cite{Straube-etal:JPCL2021} (\cref{fig:sketch}a). The core of the reaction scheme consists of two reactions that occur within the reaction compartment:
\begin{subequations}\label{4sm}
	\begin{gather}
	\ce{S ->[{k_\mathrm{cat}}][\text{\scriptsize\sf \:urease\;}] 2 P}\,, \label{4sm-R1} \\
	\ce{P + H+ <=>[k_{2}][k_{2r}] PH+}\,. \label{4sm-R2}
	\end{gather}
\end{subequations}
Reaction \eqref{4sm-R1} describes the enzyme-assisted hydrolysis of urea, \ce{CO(NH2)2}, into ammonia, \ce{NH3}, in the following denoted as substrate \ce{S} and product \ce{P}, respectively. Reaction \eqref{4sm-R2} accounts for the acidity of the medium and involves reversible conversion between the product \ce{P} and its ion form \ce{PH+} (ammonium) with the corresponding rates \cite{Eigen:ACIE1964,Hu-etal:JPCB2010, Bansagi:JPCB2014} $k_2=\SI{4.3e10}{M\tothe{-1}\s\tothe{-1}}$ and $k_{2r}=\SI{24}{\s\tothe{-1}}$.
The effective speed $v([\ce{S}],[\ce{H+}])= k_\mathrm{cat}([\ce{S}],[\ce{H+}]) [\ce{S}]$ of reaction \eqref{4sm-R1} depends on the concentrations [\ce{S}] and [\ce{H+}] of substrate and protons, respectively; or equivalently, on the level of $\ce{pH}=-\log_{10} ([\ce{H+}] / \SI{1}{M})$ and is given by the effective rate\cite{Alberty-Massey:BBA1954,Fidaleo-Lavecchia:CBEQ2003, Bansagi:JPCB2014}
\begin{align}
k_\mathrm{cat}([\ce{S}],[\ce{H+}])
& = k_\mathrm{cat}^\mathrm{M}([\ce{S}]) f_\mathrm{H}([\ce{H+}])\,. \label{eq:kcat}
\end{align}
The first factor describes the dependence on the substrate as captured by the Michaelis--Menten kinetics,
\begin{align}
k_\mathrm{cat}^\mathrm{M}([\ce{S}]) & = \frac{v_\mathrm{max} } {
	K_\mathrm{M}+[\ce{S}]} \label{eq:kcat-MM}
\end{align}
with the Michaelis--Menten constant\cite{Krajewska:JMCBE2009,Hu-etal:JPCB2010, Bansagi:JPCB2014} $K_\mathrm{M}=\SI{3e-3}{M}$.
This implies that the reaction speed $v$ grows linearly with [\ce{S}] at small $[\ce{S}] \ll K_\mathrm{M}$ and monotonically saturates at its maximum value $v_\mathrm{max}$ that would be attained in the absence of \ce{pH} effects.
The second factor implements the symmetric bell-shaped dependence of the reaction speed on the acidity (\cref{fig:sketch}c, note the logarithmic scale):
\begin{align}
f_\mathrm{H}([\ce{H+}])
& = \frac{1}{1 +  [\ce{H+}]/K_\mathrm{E1} +K_\mathrm{E2} / [\ce{H+}]}  \,, \label{eq:fH}
\end{align}
which attains its maximum value $\max_{[\ce{H+}]} f_\mathrm{H}([\ce{H+}]) = (1+2\sqrt{K_\mathrm{E2} / K_\mathrm{E1}})^{-1}$ at the hydrogen ion concentration $[\ce{H+}] = \sqrt{K_\mathrm{E1} K_\mathrm{E2}}$. For the constants chosen as \cite{Krajewska:JMCBE2009,Hu-etal:JPCB2010, Bansagi:JPCB2014} $K_\mathrm{E1}=\SI{5e-6}{M}$ and $K_\mathrm{E2}=\SI{2e-9}{M}$, it implies that the speed of reaction is maximum at the normal value of $\ce{pH}=7$, but is strongly suppressed when shifted from this optimal value to the regions of lower (acid) or higher (basic) \ce{pH}.

The core reactions \eqref{4sm-R1} and \eqref{4sm-R2} are accompanied by the exchange with a reservoir and the decay of products (\cref{fig:sketch}d); the reservoir acts as a buffer of substrate and pH, originally expressed by the reactions
\ce{S <=>[k_\mathrm{S}] {\ce{S}_\mathrm{ext}}} and
\ce{H+ <=>[k_\mathrm{H}] {\ce{H+}_\mathrm{ext}}}.
This corresponds to the setting of the spatiotemporal master equation\cite{winkelmann2016spatiotemporal,winkelmann2021mathematical}; in particular, it relies on well-mixed conditions within the vesicle and in the reservoir and it neglects possible non-Markovian effects in the transport through the membrane \cite{froemberg2021generalized,hansen2013anomalous}.
By assuming a sufficiently large reservoir such that the amounts of $\ce{S}_\mathrm{ext}$ and $\ce{H+}_\mathrm{ext}$ are changed only marginally, we consider the reservoir concentrations as fixed
values $[\ce{S}_\mathrm{ext}]$ and $[\ce{H+}_\mathrm{ext}]$. The exchange reactions are then effectively replaced by
\begin{align}\label{4sm-exch}
\ce{S <=>[k_\mathrm{S}][{k_\mathrm{S} [\ce{S}_\mathrm{ext}]}] \varnothing}
\qquad \text{and} \qquad
\ce{H+ <=>[k_\mathrm{H}][k_\mathrm{H} {[\ce{H+}_\mathrm{ext}]}] \varnothing} \,.
\end{align}
The formulation of the reaction scheme is completed by specifying the decay of products or their outflow out of the reaction compartment by the reactions
\begin{align}\label{4sm-out}
	\ce{P ->[k] \varnothing}
  \qquad \text{and} \qquad
	\ce{PH+ ->[k] \varnothing }\,.
\end{align}

The set of reaction rate equations that corresponds to reactions \eqref{4sm}, \eqref{4sm-exch} and \eqref{4sm-out} reads:
\begin{subequations} \label{eq:rre4}
	\begin{align}
	\frac{d[\ce{S}]}{dt} & = -k_\mathrm{cat}([\ce{S}],[\ce{H+}])[\ce{S}]+k_\mathrm{S}([\ce{S}_\mathrm{ext}]-[\ce{S}])\,,  \label{eq:rre4-s} \\
	\frac{d[\ce{H+}]}{dt} & = k_{2r}[\ce{PH+}]-k_2[\ce{P}][\ce{H+}] + k_\mathrm{H}([\ce{H+}_\mathrm{ext}]-[\ce{H+}])\,, \label{eq:rre4-h+} \\
	\frac{d[\ce{P}]}{dt} & = 2 k_\mathrm{cat}([\ce{S}],[\ce{H+}])[\ce{S}] + k_{2r}[\ce{PH+}]-k_2[\ce{P}][\ce{H+}]-k [\ce{P}]\,,  \label{eq:rre4-p} \\
	\frac{d[\ce{PH+}]}{dt} & = k_2[\ce{P}][\ce{H+}] -  k_{2r}[\ce{PH+}] - k [\ce{PH+}]\,, \label{eq:rre4-ph+}
	\end{align}
\end{subequations}
which we will refer to as four-variable model in the following.

Focusing on the oscillatory regime, we stick to the parameter values used previously \cite{Straube-etal:JPCL2021}. Thus, the rates of urea and proton transport correspond to $k_\mathrm{S} = \SI{1.4e-3}{\s\tothe{-1}}$ and $k_\mathrm{H} =\SI{9e-3}{\s\tothe{-1}}$, respectively; the outflow rates of both products are set to $k = k_\mathrm{S}$. For the maximum speed we use the value $v_\mathrm{max} =\SI{1.85e-4}{M\, \s^{-1}}$, which corresponds to an urea concentration of \SI{50}{U}. The external concentrations are fixed to $[\ce{S}_\mathrm{ext}]=\SI{3.8e-4}{M}$ and $[\ce{H+}_\mathrm{ext}]=\SI{1.3e-4}{M}$ and the initial concentrations inside the vesicle
are $[\ce{S}]_0 =\SI{5e-5}{M}$ and $[\ce{H+}]_0=\SI{e-5}{M}$.

\begin{figure}[t]
	\centering
	\includegraphics[width=0.55\textwidth]{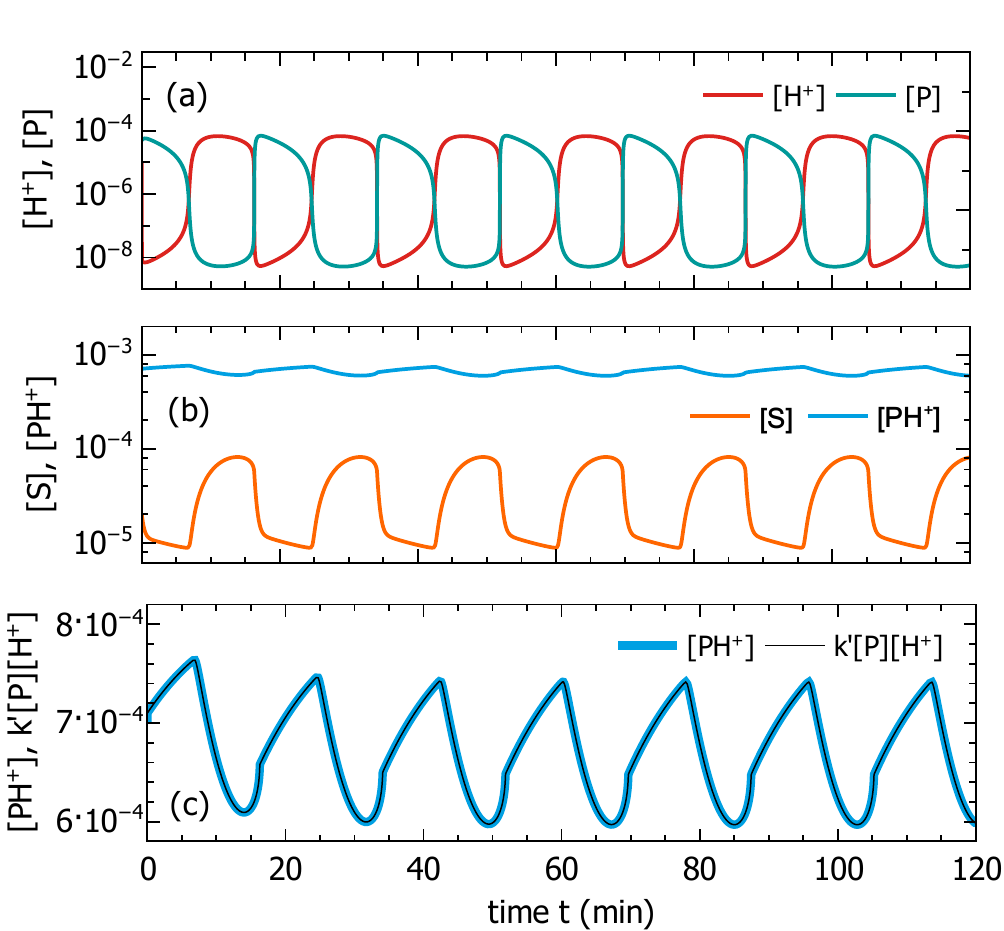}
	\caption{Evolution of the concentrations of [\ce{H+}], [\ce{P}], [\ce{S}], and [\ce{PH+}] in \si{M} from the numerical solution of the four-variable model, \cref{eq:rre4}, on logarithmic scales (panels (a) and (b)).
    Panel (c) shows the behaviour of [\ce{PH+}] (blue line) on linear scales, which coincides with the combination $k'[\ce{P}][\ce{H+}]$ (thin black line), see \cref{eq:qssa-ph+}.}
    \label{fig:4-var-model}
\end{figure}

For these parameters, the four-variable model [\cref{eq:rre4}] shows oscillatory behavior in the concentrations $[\ce{S}]$, $[\ce{H+}]$, $[\ce{P}]$ and $[\ce{PH+}]$ as exemplified in \cref{fig:4-var-model}; note the logarithmic scale in panels (a) and (b).
This evolution of the concentrations essentially reproduces that of the corresponding molecular populations in the large-vesicle limit, reported in Fig.~2 of Ref.~\citenum{Straube-etal:JPCL2021}; the periodic variation of the \ce{pH} level, roughly between $3.5$ and $8.5$, reflects the behavior of [\ce{H+}] and is shown in \cref{fig:sketch}f. The concentrations of $\ce{H+}$ and $\ce{P}$ oscillate in anti-phase over four orders of magnitude, and the evolution of $[\ce{S}]$ and $[\ce{PH+}]$ shows the same periodic behavior, but with a much smaller amplitude, \cref{fig:4-var-model}b.

\section{Quasi-steady state approximation for \ce{PH+}}

Aiming at a characterization of the limit cycle of the urea--urease oscillator, we will first pursue a dimensional reduction of the dynamical system by elimination of inessential variables. Then, we will identify the actual degrees of freedom of the system and show that the dynamical system is effectively a two-dimensional one.
In our previous work \cite{Straube-etal:JPCL2021}, starting from the four-variable model [\cref{eq:rre4}], we applied the quasi-steady-state approximation (QSSA) simultaneously to the variables $[\ce{PH+}]$ and $[\ce{P}]$ in an \emph{ad hoc} fashion.
Here, we follow a more general and systematic approach, yielding more accurate reduced models and, particularly, also their regimes of validity.
The model reduction occurs in two steps: In the first step, we eliminate $[\ce{PH+}]$, which leads to a three-variable model;
this step is common to all models considered below.
In the second step, we aim at a further reduction to two variables by eliminating $[\ce{P}]$, which can be performed in different ways leading to distinct models.

\subsection{Reduction to three variables}

By inspection of \cref{fig:4-var-model}, we can draw two important conclusions. First, the concentration of the ion form of the product is larger than all other concentrations, $[\ce{PH+}] \gg  [\ce{S}], [\ce{H+}], [\ce{P}]$. Second, the relative variation of $[\ce{PH+}]$ is smaller than those of the other concentrations, which suggests to approximate $[\ce{PH+}](t)$ by a constant.
However, enforcing $[\ce{PH+}](t) = \mathrm{const}$ in \cref{eq:rre4}, and thus $[\ce{P}](t)\, [\ce{H+}](t) = \mathrm{const}$, would turn the pH oscillations into a simple relaxation [\cref{eq:rre4-h+}].
Instead, we observe that there exist well-separated timescales \cite{Segel-Slemrod:SIAM-Rev1989, Wechselberger2020}:
the fast species $\ce{PH+}$ adjusts quickly, on the scale of $(k_{2r} + k)^{-1} = \SI{4.2}{ms}$, to the slowly evolving concentrations of $\ce{P}$ and $\ce{H+}$, which vary on the scale of several minutes.
This justifies to perform a QSSA for $\ce{PH+}$ by rearranging \cref{eq:rre4-ph+} into
\begin{equation}
 	[\ce{PH+}] = -(k_{2r} + k)^{-1}\,\frac{d[\ce{PH+}]}{dt} + k' [\ce{P}][\ce{H+}] \,, \qquad k' = \frac{k_2}{k_{2r}+k};
\end{equation}
and neglecting the first term on the r.h.s., which yields the relation
\begin{equation}
  [\ce{PH+}](t) = k' [\ce{P}](t)\,[\ce{H+}](t)
  \label{eq:qssa-ph+}
\end{equation}
with $k' = \SI{1.8e9}{M\tothe{-1}}$ for the rates given above.
The high accuracy of this approximation is corroborated by \cref{fig:4-var-model}c,
which shows that the temporal behavior of both sides of \cref{eq:qssa-ph+} coincides at all times, although none of the involved concentrations remains constant. Making use of \cref{eq:qssa-ph+} in \cref{eq:rre4-h+,eq:rre4-p}, we arrive at a reduced model that employs only three variables:
\begin{subequations} \label{eq:rre3}
	\begin{align}
	\frac{d[\ce{S}]}{dt} & = -k_\mathrm{cat}([\ce{S}],[\ce{H+}])[\ce{S}]+k_\mathrm{S}([\ce{S}_\mathrm{ext}]-[\ce{S}])\,,  \label{eq:rre3-s} \\
	\frac{d[\ce{H+}]}{dt} & = -kk'[\ce{P}][\ce{H+}] + k_\mathrm{H}([\ce{H+}_\mathrm{ext}]-[\ce{H+}])\,, \label{eq:rre3-h+} \\
	\frac{d[\ce{P}]}{dt} & = 2 k_\mathrm{cat}([\ce{S}],[\ce{H+}])[\ce{S}] - kk'[\ce{P}][\ce{H+}]-k[\ce{P}]\,.  \label{eq:rre3-p}
%	\frac{d[\ce{P}]}{dt} & = 2 k_\mathrm{cat}([\ce{S}],[\ce{H+}])[\ce{S}] - k [\ce{P}] (k'[\ce{H+}]+1)\,, \label{eq:rre3-p}
	\end{align}
\end{subequations}
Note that while simplifying \cref{eq:rre3-p}, in accord with the above reasoning, we could have neglected the last term, $-k[\ce{P}]$. Indeed, the assumption $[\ce{PH+}] \gg [\ce{P}]$ together with \cref{eq:qssa-ph+} is equivalent to the requirement $k' [\ce{H+}] \gg 1$.
As independently confirmed by a previous study \cite{Bansagi:JPCB2014}, this is a reasonable simplification for modeling pH oscillations. However, to analyze the dynamics in the whole phase plane and to keep the predictions of the reduced models as close as possible to those of the original four-variable model, we retain this term.

\begin{figure}[t]
	\centering
	\includegraphics[width=0.55\textwidth]{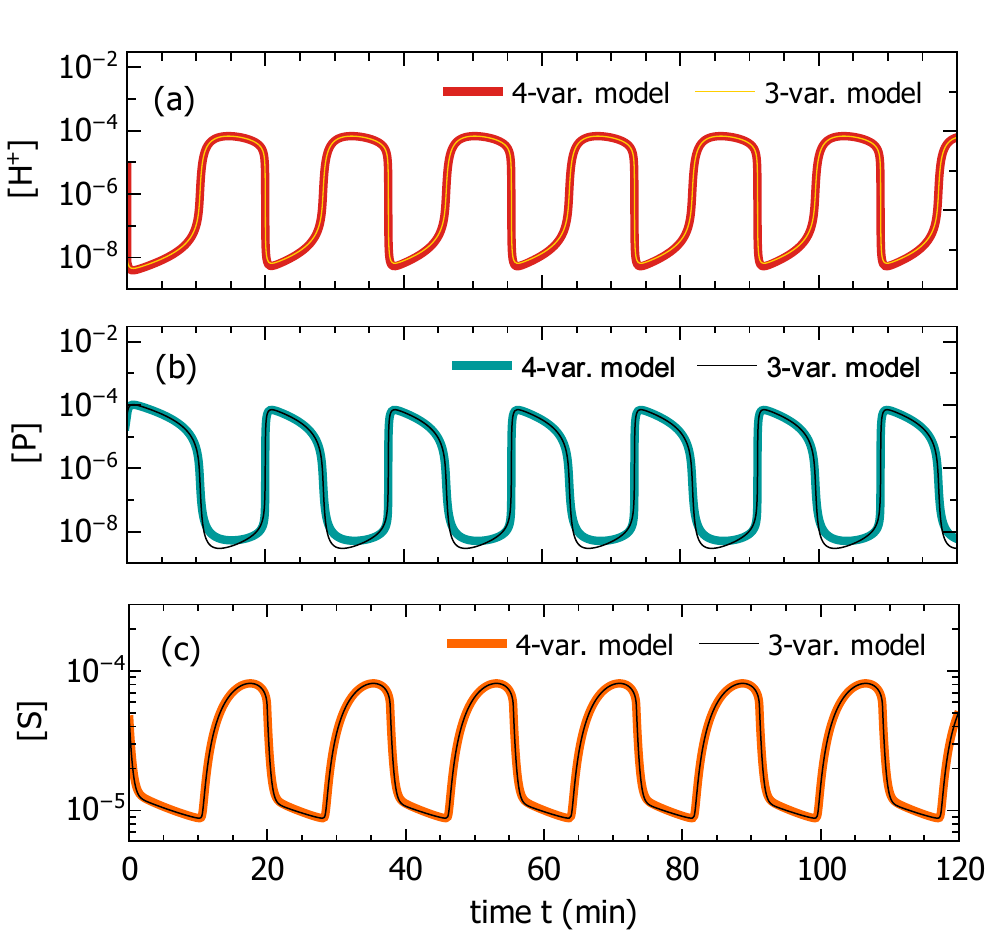}
	\caption{Comparison of the solutions of the four- and three-variable models given by \cref{eq:rre4,eq:rre3}, respectively. The panels show the evolution of the concentrations of (a) acid [\ce{H+}], (b) product [\ce{P}], (c) and substrate [\ce{S}].
	The comparison of the corresponding \ce{pH} levels is shown in \cref{fig:sketch}f.}
	\label{fig:3-and-4-var-model}
\end{figure}

The dynamical system in \cref{eq:rre3} can be interpreted as the reaction rate equations of the following effective system of in-volume reactions (\cref{fig:sketch}e):
\begin{equation}
	\ce{S ->[{k_\mathrm{cat}}][\text{\scriptsize\sf \:urease\;}] 2 P}\,, \qquad
%	\ce{P + H+ <=>[k_2][k_{2r}] \varnothing} \,.
    \ce{P + H+ ->[kk'] \varnothing}\,, %\qquad \ce{P ->[k] \varnothing}
	\label{3sm}
\end{equation}
amended by the exchange reactions \eqref{4sm-exch} of \ce{S} and \ce{H+} with the reservoir and the decay of \ce{P}, see the first reaction in \cref{4sm-out} (\cref{fig:sketch}d). Thus, the product \ce{P} has two channels to escape from the vesicle: directly and after protonation with an effective rate. In the original reaction scheme involving four species, the second channel exists indirectly, via escape of \ce{PH+}.

We also note that despite relation \eqref{eq:qssa-ph+} is fulfilled with high accuracy, the evolutions described by the three- and four-variable models are not identical (Figs.~\ref{fig:sketch}f and \ref{fig:3-and-4-var-model}). Slight quantitative differences in the predictions of the two models are visible for \ce{P} and \ce{pH} (and, equivalently, for $[\ce{H+}]$). Interestingly, whereas the deviations are tiny for \ce{pH}, they are more pronounced for \ce{P}, yet eventually less important because we are generally not interested in the dynamics of the product. What is essential is that the reduced system \eqref{eq:rre3} preserves not only all qualitative features of the four-variable model, but it remains quantitatively reliable in predicting the period of the oscillations. Thus, the three-variable model \eqref{eq:rre3} serves as a highly accurate approximation.

\subsection{Reduction to two variables}

The further reduction of the three-variable model to two species by means of a QSSA for $[\ce{P}]$ appears less justified than the elimination of $[\ce{PH+}]$, as we will see below. Instead, we shall put forward a solution that yields an essentially exact reduction.
It combines a scale separation argument and the condition in \cref{eq:qssa-ph+}.
As noted above, $[\ce{PH+}](t)$ oscillates slowly and with a relative amplitude of about 10\% (\cref{fig:4-var-model}c). In contrast, $[\ce{P}](t)$ and $[\ce{H+}](t)$ vary over four orders of magnitude on the same time scale (\cref{fig:3-and-4-var-model}) and their logarithms oscillate very similarly, but in antiphase (\cref{fig:4-var-model}a).
Taking time derivatives, these observations suggest that the r.h.s. of the relation
\begin{equation}
  \frac{d}{dt}\log([\ce{P}](t)) + \frac{d}{dt}\log([\ce{H+}](t)) = \frac{d}{dt} \log([\ce{PH+}](t)])
\end{equation}
is negligible, which implies the approximate, yet accurate constraint
\begin{equation}
  [\ce{P}](t)\,[\ce{H+}](t) \approx \mathrm{const}
  \label{eq:p-h-constraint}
\end{equation}
and thus a tight coupling of the dynamics of $[\ce{P}](t)$ and $[\ce{H+}](t)$.

\begin{figure}[t] \centering
 \includegraphics{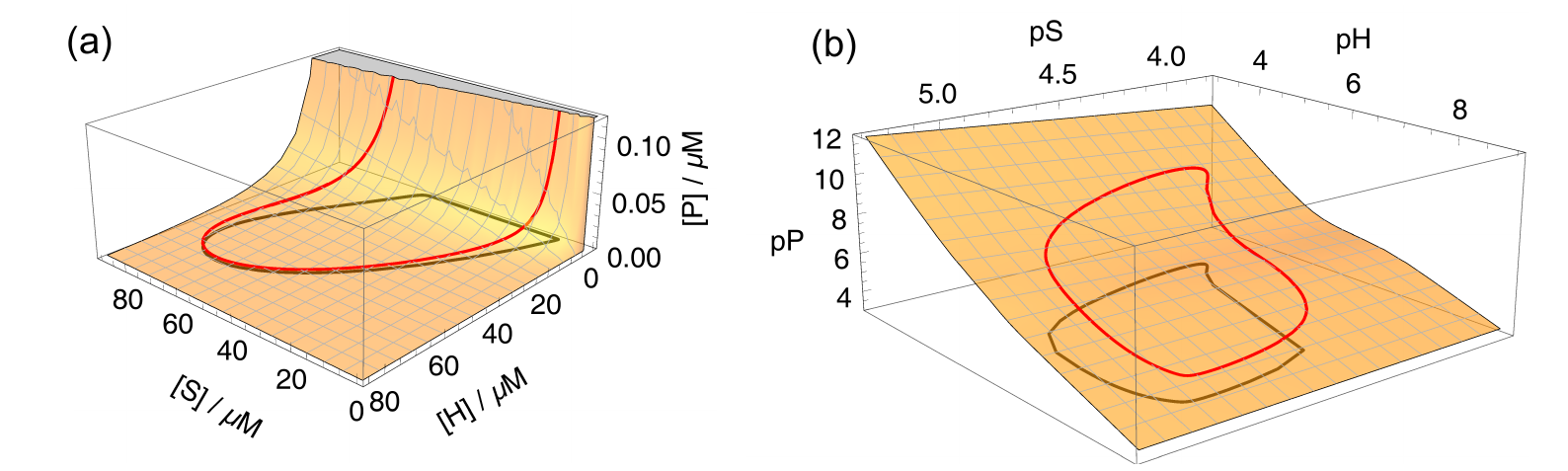}
 \caption{Phase plot of the limit cycle (red) of the three-variable model, \cref{eq:rre3}, using linear (a) and logarithmic (b) scales. The surface is given by \cref{eq:p}, which represents the constraint, \cref{eq:p-h-constraint}. The black line is the orthogonal projection of the limit cycle to the plane $\ce{pP} = 4$ and $[\ce{P}] = 0$, respectively. In panel~(b), we use the symbols $\ce{pX}=-\log_{10}([\ce{X}]/\SI{1}{M})$ for $X=\ce{S}, \ce{H+},\ce{P}$; note that small values of, e.g., \ce{pS} correspond to large concentrations $[\ce{S}]$.}
 \label{fig:phaseplot3d}
\end{figure}

\Cref{fig:phaseplot3d} provides a visualization of the constraint in a three-dimensional phase plot of the three-variable model \eqref{eq:rre3}.
First, we note that the structure of the limit cycle is not well resolved on linear scales (panel~(a)),
especially at small concentrations of $[\ce{H+}]$ (high \ce{pH}).
However, utilizing logarithmic scales, the structure of the limit cycle is uniformly well exhibited in the whole range of values (panels~b).
This observation is another indication that the oscillator studied here behaves differently than the conventional examples of \ce{pH} oscillators \cite{Orban:ACR2015}.
Second, one observes clearly that the dynamics in the three variables $[\ce{S}]$, $[\ce{H+}]$, and $[\ce{P}]$ is tightly confined to a two-dimensional manifold of roughly hyperbolic shape in the linear representation, see \cref{eq:p-h-constraint}. The constraining manifold simplifies approximately to a plane in the logarithmic representation, which is tilted against all three coordinate axes; in particular, it deviates strongly from any plane given by a constant value of $[\ce{P}]$. This fact signifies clearly that the naive orthogonal projection of the phase flow to such a plane for the elimination of the variable $[\ce{P}]$ would be a poor approximation of the true dynamics; an issue we will expand on below (Section~``Failure of QSSA for the product'').

Equipped with these insights, we proceed to eliminate the variable $\ce{P}$ from the three-dimensional model.
For clarity of the subsequent analysis, we switch to dimensionless variables $s(t):=[\ce{S}](t)/[\ce{S}_\textrm{ext}]$, $h(t):=[\ce{H+}](t) / [\ce{H+}_\textrm{ext}]$, and $p(t):=k'\, [\ce{P}](t)$, which turns \cref{eq:rre3} into
\begin{subequations} \label{eq:rre3nondim}
	\begin{align}
	\frac{ds}{dt} & = -k_\mathrm{cat}(s,h)\, s + k_\mathrm{S} \,(1 - s)\,,  \label{eq:rre3nondim-s} \\
	\frac{dh}{dt} & = -k \,p \,h + k_\mathrm{H}\,(1 - h)\,, \label{eq:rre3nondim-h+} \\
	\frac{dp}{dt} & = 2\, \alpha_{\textrm{S}}\, k_\mathrm{cat}(s,h)\, s - \alpha_{\textrm{H}}\, k\, p\, h\,  u(h)\,,  \label{eq:rre3nondim-p}
	\end{align}
\end{subequations}
where $\alpha_\textrm{S}:=k'[\ce{S}_\textrm{ext}]$, $\alpha_\textrm{H}:=k'[\ce{H+}_\textrm{ext}]$, and $u(h):=1+(\alpha_{\textrm{H}} h)^{-1}$.
In terms of these variables, we rewrite \cref{eq:p-h-constraint} as
\begin{equation}
  p(t)\frac{dh(t)}{dt} + \frac{dp(t)}{dt}h(t) \approx 0\,.
  \label{eq:elimcond}
\end{equation}
After substitution of the time derivatives using \cref{eq:rre3nondim-h+,eq:rre3nondim-p},
the constraint assumes the form of a quadratic equation in $p=p(s,h)$,
\begin{align}
	p^2 + b(h)\,p - c(s,h) =0\,, 
	\label{eq:quadeq-p}
\end{align}
with the dimensionless coefficients
\begin{align}
	b(h) &= \alpha_{\textrm{H}}\, h\,  u(h) + ( 1 - h^{-1} ) \, k_\mathrm{H} /k \label{eq:cffs-b}
\intertext{and}
	c(s,h) &= \frac{2\, \alpha_{\textrm{S}}\, k_\mathrm{cat}(s,h)}{k}s \ge 0\,. \label{eq:cffs-c}
\end{align}
\Cref{eq:quadeq-p} possesses two roots, $p_{\pm}(s,h)$,
%$p_{\pm}=( -b \pm \sqrt{b^2+4c} ) / 2$,
and selecting the positive solution, $p(s,h) \ge 0$ for all $s, h \geq 0$, we obtain
\begin{align}
p(s,h) = \frac{b(h)}{2} + \frac{1}{2}\sqrt{b(h)^2+4c(s,h)}\,. \label{eq:p}
\end{align}
In particular, the time evolution of the rescaled concentration $[\ce{P}]$ of products is enslaved to the evolution of $[\ce{S}]$ and $[\ce{H+}]$ and is given by $p(t) = p(s(t), h(t))$.

This solution for $p(t)$ allows us to eliminate $p(t)$ as a variable from the three-variable model in \cref{eq:rre3nondim}. In particular, one verifies that \cref{eq:rre3nondim-p} is automatically satisfied and can be dropped.
The remaining \cref{eq:rre3nondim-s,eq:rre3nondim-h+} yield the two-variable model:
\begin{subequations} \label{eq:rre2}
	\begin{align}
	\frac{ds}{dt} & = F(s,h) := -k_\mathrm{cat}(s,h) \, s + k_\mathrm{S}\, (1-s)\,,  \label{eq:rre2-s} \\
	\frac{d h}{dt} & = G(s,h) := -k \,p(s,h) \,h + k_\mathrm{H}\,(1-h)\,, \label{eq:rre2-h+}
	\end{align}
\end{subequations}
where $p(s,h)$ is defined by relation \eqref{eq:p} with the coefficients $b(s,h)$ and $c(s,h)$ given by \cref{eq:cffs-b,eq:cffs-c}.
We stress that \cref{eq:p} is
a consequence of \cref{eq:p-h-constraint} or, equivalently,  \cref{eq:elimcond}.
The two-variable model [\cref{eq:rre2} with \cref{eq:p}] is thus a virtually exact representation of the three-variable model \eqref{eq:rre3}.
However, differently from the latter, \cref{eq:rre2} does not have a meaningful interpretation as rate equations of a system of effective reactions unless one accepts $k\, p(s,h)$ as an effective decay rate of $\ce{H+}$. Since the introduction of nonelementary rates as a result of reduction on the deterministic level may lead to significant quantitative and even qualitative errors in stochastic simulations,\cite{Thomas-etal:JCP2010, Thomas-etal:BMCSB2012} it is favorable to use the three-variable alternative as a stochastic model.

\section{Limit cycle and structure of the phase flow}

\subsection{Nullclines}

General properties of the limit cycle can be understood from the geometric structure of the phase flow $(F(s,h), G(s,h))$  of a dynamical system. Helpful characteristics specifying the structure of the phase flow or phase portrait are nullclines. For the two-variable model \eqref{eq:rre2}, the $\ce{S}$ and $\ce{H+}$ nullclines are defined by the conditions $F(s,h)=0$ and $G(s,h)=0$, respectively. By construction, on a given nullcline the flow is perpendicular to the axis of the corresponding variable.  The intersection of all nullclines defines all fixed points of the phase flow, which represent the steady state solutions of \cref{eq:rre2}, i.e., the points where all time derivatives vanish. For the present two-variable model, both nullclines can be obtained analytically.

The $\ce{S}$ nullcline is calculated by putting $F(s,h)=0$ in \cref{eq:rre2-s}, which on account of \cref{eq:kcat} leads to a quadratic equation in $s$:
\begin{equation}
  s^2+\beta(h)s-\bar K_\mathrm{M}=0
  \quad \text{with} \quad
  \beta(h) := \bar K_\mathrm{M} - 1 + \bar v_\mathrm{max} f_\mathrm{H}(h) \,,
\end{equation}
where $\bar v_\mathrm{max}:=v_\mathrm{max}/(k_\mathrm{S} [\ce{S}_\mathrm{ext}])$, 
$\bar K_\mathrm{M} := K_\mathrm{M}/[\ce{S}_\mathrm{ext}]$ and $f_\mathrm{H}(h)$ is given by \cref{eq:fH}.  Out of the two roots $s_{\pm}(h)$, only the one describing non-negative substrate concentration is physically relevant, thus yielding the $\ce{S}$ nullcline,
\begin{align}
s_\mathrm{S}^\mathrm{nc}(h) = s_+(h) :=
 -\frac{\beta(h)}{2} + \frac{1}{2}\sqrt{\beta(h)^2+4 \bar K_\mathrm{M}}\,.
\label{eq:nullcline-s}
\end{align}
Note that the dependence of $s_\mathrm{S}^\mathrm{nc}(h)$ on $h$ inherits its shape directly from \cref{eq:fH}. Its maximum is located at $h_\textrm{m}=\sqrt{K_\textrm{E1} K_\textrm{E2}}/[\ce{H+}_\textrm{ext}]$ as determined from the condition $s_+'(h_m) = 0$, which is found to be equivalent to $f_\textrm{H}'(h_\textrm{m} [\ce{H+}_\textrm{ext}])=0$.

To obtain the \ce{H+} nullcline, we set $G(s,h)=0$ in \cref{eq:rre2-h+} and, using the definition of $p(s,h)$, after some tedious algebra, we arrive at the intermediate relation,
\begin{equation}
  2 \kappa k_\mathrm{cat}(s,h)  s
  =k_\mathrm{H}(1-h) u(h)
\end{equation}
with $\kappa=\alpha_{\textrm{S}}/\alpha_{\textrm{H}}=[\ce{S}_\mathrm{ext}]/[\ce{H+}_\mathrm{ext}]$.
More straightforwardly, we can derive this result by making use of \cref{eq:elimcond}, which implies that $dh/dt=0$ is equivalent to $dp/dt=0$ for $p(t), h(t)>0$:
to this end, we set to zero the time derivative in \cref{eq:rre3-p}, solve for
\begin{equation}
k p(s,h) h = \frac{2 \kappa k_\mathrm{cat}(s,h) }{u(h)} \,s \,, \label{eq:kph}
\end{equation}
and substitute it in \cref{eq:rre2-h+}. Finally, inserting the definition of $k_\mathrm{cat}(s,h)$, \cref{eq:kcat}, into the intermediate equation for $s$ leads us to the \ce{H+} nullcline,
\begin{align}
s_\mathrm{H}^\mathrm{nc}(h)=
\bar K_\mathrm{M} \left[\frac{2 \kappa k_\textrm{S} \bar v_\mathrm{max}f_\mathrm{H}(h)} {k_\mathrm{H} (1-h)u(h)}-1\right]^{-1}
\,.
\label{eq:nullcline-h}
\end{align}

\subsection{Phase flow}

\begin{figure}
	\centering
	\includegraphics[width=0.9\textwidth]{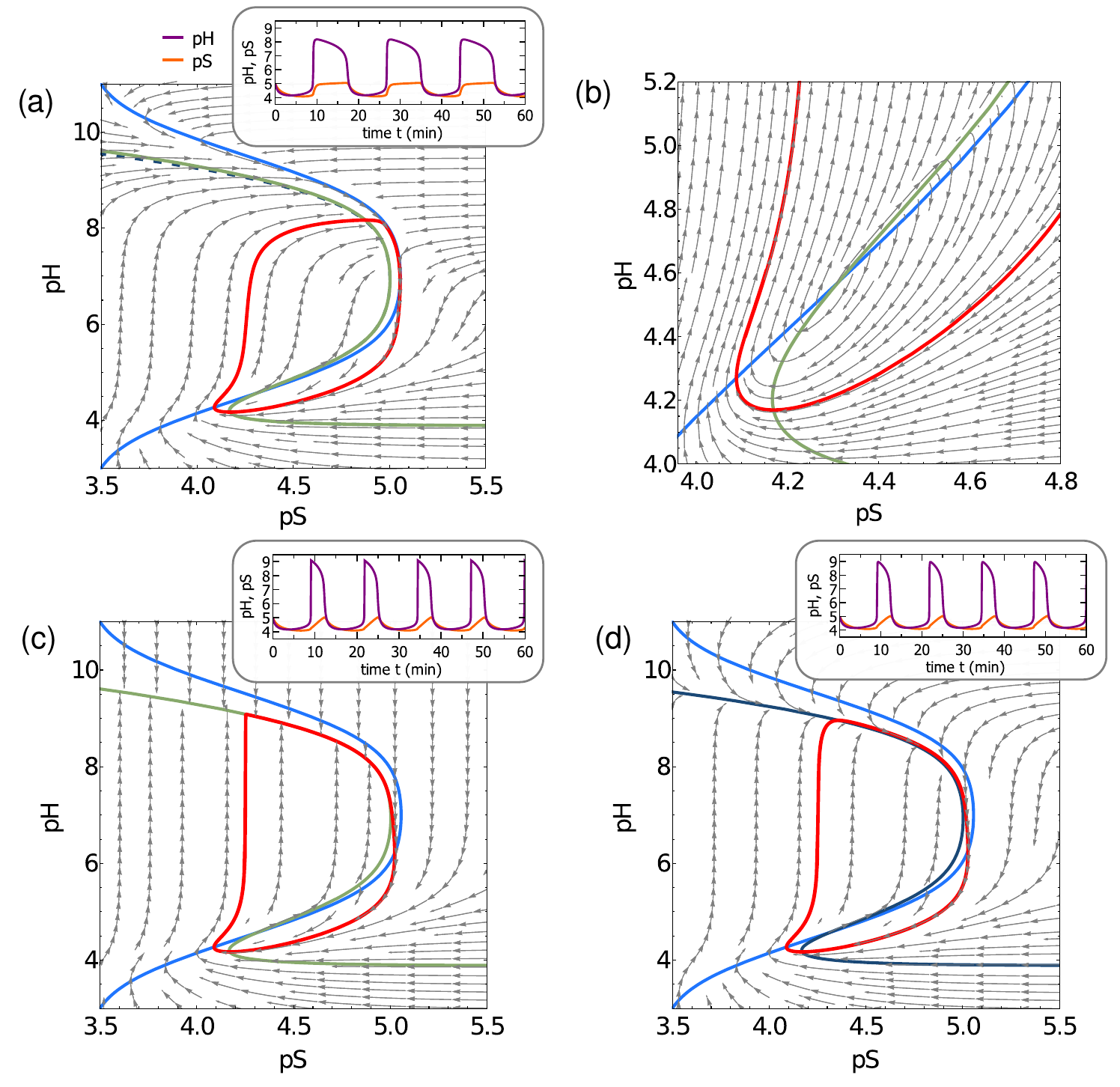}
	\caption{Phase flows (grey arrows) and limit cycles (red lines) in the \ce{pS}-\ce{pH} plane for the two-variable models obtained from two different reduction schemes of the three-variable model [\cref{eq:rre3}]:
	(a)~~accurate model [\cref{eq:rre2}] from the exact elimination of the product $p$;
	(b)~~close-up of the vicinity of the fixed point at $(\ce{pS}_*, \ce{pH}_*) \approx (4.31,4.57)$ shown in (a);
	(c)~~model in \cref{eq:rre2A} from imposing the QSSA for $p(t)$;
	(d)~~model in \cref{eq:rre2B} from explicitly accounting for the water ionization, \cref{eq:rre3A}, and then imposing the QSSA on the product, this model corresponds to the two-variable model studied by \citet{Bansagi:JPCB2014}.
	The \ce{S} (blue lines) and \ce{H+} (green lines) nullclines are the same for the models in (a)--(c) and are given by \cref{eq:nullcline-s,eq:nullcline-h}, respectively.
	Including the water ionization leads to a slight shift of the \ce{H+} nullcline from the green line to the dashed blue line in (a) and to the dark blue line in (d), \cref{eq:nullcline-h-B}.
	}
	\label{fig:phase-plot}
\end{figure}

The origin of the limit cycle is easily understood from the phase portrait of the two-dimensional dynamical system, \cref{eq:rre2}, and the associated nullclines, \cref{eq:nullcline-s,eq:nullcline-h}. As mentioned earlier, a remarkable feature of this system is that the structure of the phase flow is best unveiled on logarithmic scales---in contrast to conventional examples of oscillators. Therefore, instead of the original variables $s$ and $h$, we will use $\ce{pS}$ and $\ce{pH}$ as axes of the phase plane.
The phase flow of the two-variable model together with its nullclines and the limit cycle are shown in \cref{fig:phase-plot}a. For the parameters considered here, the nullclines intersect only in a single, repelling fixed point enclosed by the limit cycle. 
The limit cycle was obtained from the numerical solution of $ds/dt=F(s,h)$ and $dh/dt=G(s,h)$ after a sufficiently long initial relaxation time.

Qualitatively, the shape of the limit cycle is determined by the nullclines and the two-dimensional flow field $(F(s,h),G(s,h))$.
The \ce{S} nullcline (\cref{eq:nullcline-s}, blue line in \cref{fig:phase-plot}a) has a rotated bell shape, inherited from the function $f_\textrm{H}([\ce{H+}])$ (see \cref{eq:fH} and \cref{fig:sketch}f); and the \ce{H+} nullcline (\cref{eq:nullcline-h}, green line in \cref{fig:phase-plot}a) is of a reversed s-like shape. The requirement on the flow field that, at any point of the nullclines, there is no motion along the corresponding direction means that the flow on the \ce{S} nullcline can only point along the \ce{pH} direction (vertical arrows) and on the \ce{H+} nullcline along the \ce{pS} direction (horizontal arrows).
In the high \ce{pH} regime, both nullclines run closely together such that the \ce{H+} nullcline pushes the flow towards the \ce{S} nullcline. This causes a channeling of the flow between the two nullclines towards the apex of the \ce{S} nullcline, $(\ce{pS}, \ce{pH}) \approx (5,7)$, where $s_\mathrm{S}^\mathrm{nc}(h)$ attains its maximum value. There, the flow points downwards, along the \ce{pH} axis, and is tightly restricted with respect to \ce{pS} (i.e., horizontally), concomitantly the \ce{pH} value drops rapidly. All phase trajectories, irrespective of their starting point, eventually approach this apex point arbitrarily closely and follow the \ce{S} nullcline for a moment, which essentially defines a piece of the limiting trajectory (limit cycle, red line).
After this point, the \ce{S} nullcline bends away from the vertical, but the trajectory keeps following the flow field and revolves around the fixed point until the orbit closes, which forms the limit cycle. Thus, the limit cycle is determined as the trajectory that starts in the apex of the \ce{S} nullcline, where the curve attains its largest \ce{pS} value.

The \ce{H+} and \ce{S} nullclines, shown in \cref{fig:phase-plot}a, have a single intersection point, defining a fixed point, which is enclosed  by the limit cycle. The flow field in the vicinity of the fixed point (\cref{fig:phase-plot}b) indicates that this fixed point acts as a repeller. This observation is in agreement with the Poincar{\'e}--Hopf theorem of indices \cite{Guckenheimer-Holmes1983}. It states that the index $I(\gamma)$ of any closed curve $\gamma$ equals the sum of the indices $I_k$ of all the enclosed fixed points, $I(\gamma)=\sum_k I_k$, where one assigns the index $I(\gamma)=+1$ to a periodic orbit $\gamma$, while the index of a saddle fixed point is $I=-1$ and the index of a node, a spiral (focus) or a center is $I=+1$.
As a consequence, the limit cycle must enclose at least one fixed point (which must not be a saddle point).
Moreover, the existence of a single repelling fixed point as in \cref{fig:phase-plot}b automatically implies that it is either an unstable spiral or node. A more detailed stability analysis will be given below (Section~`Fixed point and its stability').

\section{Failure of QSSA for the product}

In the literature, it was suggested to apply the QSSA to the concentration $[\ce{P}]$ of products \cite{Bansagi:JPCB2014, Straube-etal:JPCL2021}. Here we elaborate on the consequences of this approximation and show that although the nullclines remain the same as in the exact reduction scheme, it qualitatively changes the phase flow and significantly affects the oscillations. In hindsight, it is clear that generally such an approximation cannot be consistent with the QSSA for $[\ce{PH+}]$, which leads to the three-variable model, \cref{3sm}, and constrains the three-dimensional flow to a two-dimensional manifold as discussed above (\cref{fig:phaseplot3d}).

This elimination scheme follows directly from \cref{eq:rre3nondim} by enforcing the QSSA for $p(t)$. As has already been shown, setting $dp/dt=0$ in \cref{eq:rre3nondim-p} yields expression \eqref{eq:kph} for the combination $k p(s,h) h$.
Substituting it in \cref{eq:rre3nondim-h+}, we arrive at the model:
\begin{subequations} \label{eq:rre2A}
	\begin{align}
	\frac{ds}{dt} & = -k_\mathrm{cat}(s,h) s + k_\mathrm{S} (1-s)\,,  \label{eq:rre2A-s} \\
	\frac{d h}{dt} & = - \frac{2 \kappa k_\mathrm{cat}(s,h)}{u(h)} \, s + k_\mathrm{H}(1-h)\,. \label{eq:rre2A-h+}
	\end{align}
\end{subequations}
In our previous study,\cite{Straube-etal:JPCL2021} this model was used with $u(h)=1$ to qualitatively obtain the structure of the phase portrait exhibited by the numerical solution of the four-variable model. As earlier, the \ce{H+} nullcline corresponding to \cref{eq:rre2A-h+} follows simply by setting $dh/dt=dp/dt=0$ in \cref{eq:rre3nondim-h+,eq:rre3nondim-p} and solving for $s(h)$. Expressing the combination $kph$ from the equation for $h$ and equating it with that from \cref{eq:kph}, we end up with the \ce{H+} nullcline identical to \cref{eq:nullcline-h}. The \ce{S} nullcline is given by \cref{eq:nullcline-s} due to the coincidence of the equations for $s$, cf.~\cref{eq:rre2-s} and \cref{eq:rre2A-s}.

By comparing the phase plots of the models given by \cref{eq:rre2} and \cref{eq:rre2A}, we immediately notice that although the nullclines of both models are the same, the flow fields in the upper half of the plots (large \ce{pH}) are drastically different (\cref{fig:phase-plot}a,~c). In particular, the behavior of the flow field on the upper branch of the \ce{H+} nullcline described by \cref{eq:rre2A} degenerates (\cref{fig:phase-plot}c). By construction, the flow on the \ce{H+} nullcline has to point in the \ce{pS} direction (horizontal arrows), meaning the absence of the \ce{pH} (vertical) component of the flow field, $dh/dt=0$; the latter has to comply with the fact that $dh/dt$ has opposite signs above and below the nullcline. Closer inspection of the flow shows that, in contrast to model \eqref{eq:rre2}, the relative role of the vertical component of the flow field appears strongly overestimated by the QSSA such that the relation $|ds/dt| \ll |dh/dt|$ holds already slightly away from the nullcline. As a result of this improper balance of the two components, the flow field bends sharply near the \ce{H+} nullcline and follows it towards larger \ce{pS} with a small velocity (\cref{fig:phase-plot}c). In other words, the \ce{H+} nullcline acts as a virtual attractor, which is a qualitatively wrong picture as follows from the comparison with the flow of the accurate model \eqref{eq:rre2} shown in \cref{fig:phase-plot}a.
\Cref{fig:phase-plot}c reveals further that the limit cycle is formed in a qualitatively different way.
In contrast to the accurate scenario, the shape of the limit cycle is now fully set by the upper branch of the \ce{H+} nullcline, which directly affects the oscillation behavior predicted by this model. The temporal oscillation patterns (insets of \cref{fig:phase-plot}a,c) exhibit drastically different shapes. In particular, the ad-hoc model \eqref{eq:rre2A} exhibits a significantly shorter relaxation of the \ce{pH} level (\cref{fig:phase-plot}c) relative to the prediction of the accurate model \eqref{eq:rre2}. As a result, the oscillation period % $\tau$
is found to be \SI{12.68}{\minute} within model \eqref{eq:rre2A}, which underestimates the reliable prediction of
\SI{17.59}{\minute} within model \eqref{eq:rre2} by nearly $30\%$.
We note further that model \eqref{eq:rre2A}, which degenerates at high \ce{pH} and fails to capture the rounded shape of the limit cycle, also overestimates the amplitude of the \ce{pH} oscillations: it yields $\max(\ce{pH})- \min(\ce{pH}) \approx 4.9$, which is to be compared to the value $4.0$ obtained within model \eqref{eq:rre2}.

We note that the model given by \cref{eq:rre2A}, and hence the QSSA for the product \ce{P}, correspond to a special limit of the model given by \cref{eq:rre2}. It follows from the exact solution for $p(s,h)$, see \cref{eq:cffs-b,eq:cffs-c,eq:p}, under two separate conditions. First, one requires that $b(h)^2 \gg 4 c(s,h)$, which permits the expansion of the square root in \cref{eq:p},
\begin{align}
  p(s,h) & \simeq -\frac{b(h)}{2} + \frac{|b(h)|}{2} \left(1+\frac{2c(s,h)}{b(h)^2}\right)  \nonumber \\
         & =
            \begin{cases}
                c(s,h) / b(h),	& \text{if } b(h) > 0\,, \\
                -b(h),   & \text{if } b(h) < 0\,.
            \end{cases} \label{eq:p-b-large-approx}
\end{align}
Second, if $\alpha_\textrm{H} h u(h) \gg (1-h^{-1})k_\mathrm{H}/k$ one can simplify $b(h) \simeq \alpha_{\textrm{H}} h  u(h) > 0$ yielding
\begin{align}
  p(s,h) = \frac{2 \kappa k_\mathrm{cat}(s,h)}{k h u(h)} s  \ge 0 \,. \label{eq:p-qssa-approx}
\end{align}
We note that this result corresponds to setting $dp/dt=0$ in \cref{eq:rre3nondim-p}.

Thus, the two assumptions leading to \cref{eq:p-qssa-approx} define the regime of validity of the QSSA and essentially require that $b(h)$ is positive and large enough. As follows from \cref{eq:p-b-large-approx}, in the opposite regime $b < 0$ and large $|b|$, the consistent approximation for $p(s,h)$ should be different. The border case, $b(h)=0$, occurs for $h =[\sqrt{(k_\textrm{H}+k)^2+4 \alpha_\textrm{H} k k_\textrm{H} } - (k_\textrm{H}+k) ] / (2 \alpha_\textrm{H}k)$. For the parameters used here, $\alpha_\textrm{H} \approx \num{2e6} \gg 1$ and $h \approx \sqrt{k_\textrm{H}/(\alpha_\textrm{H} k)}$, which corresponds to $\ce{pH}\approx 6.17$. This means that the approximation \eqref{eq:p-qssa-approx} is justified for a part of the oscillation period only, namely for $\ce{pH} \lesssim 6$ (then $h$ is large enough and $b(h) > 0$), see inset of \cref{fig:phase-plot}a. For the rest of the period, where $\ce{pH} \gtrsim 6$, the approximation \cref{eq:p-qssa-approx} is not applicable. In this regime, either the above expansion fails completely (for $\ce{pH}\approx 6$, $b^2 \ll 4c$) or $p$ is determined by a different relation than \cref{eq:p-qssa-approx} (large values of \ce{pH}, $b^2 \gg 4c$, but  $b<0$). These are the reasons that significantly restrict the validity of the QSSA for \ce{P} and thus of the whole model given by \cref{eq:rre2A}. In particular, the preceding analysis explains why the model \eqref{eq:rre2A} exhibits a degeneracy and becomes unreliable when the \ce{pH} level is neutral or basic ($\ce{pH} \gtrsim 6$).

\section{Fixed points and stability analysis}

For the characterization of the parameter regimes where oscillatory behavior is predicted, we shall determine the fixed points of the accurate model, \cref{eq:rre2}, and analyse their stability.
The problem does not admit for a simple analytic treatment in the whole \ce{pH} range.
However, noting that the fixed point giving rise to oscillations is located in the acid regime, we exploit that, for low \ce{pH} values, \cref{eq:rre2} reduces to the more tractable model in \cref{eq:rre2A}.
This approach is corroborated by \cref{fig:phase-plot}: for the parameter set used, the models \eqref{eq:rre2} and \eqref{eq:rre2A} display very similar flow fields for $\ce{pH}\lesssim 6$ (panels (a) and (c)). The \ce{S} and \ce{H+} nullclines exhibit a single intersection, which is located in the acid regime, $\ce{pH} \approx 4.6$ (panels (a) and (b)); the intersection yields the fixed point of the flow, and the surrounding flow field indicates that it is an unstable one. We will now obtain an analytic expression for the fixed point, explore its stability and develop an overall picture of the possible equilibria.

\subsection{Stability of equilibria and domain of oscillations}

The fixed points $(s_*, h_*)$ of model \eqref{eq:rre2}, and also of model \eqref{eq:rre2A}, with $p(s,h)$ given by either \cref{eq:p} or \cref{eq:p-qssa-approx}, respectively, are determined by the conditions
\begin{subequations} \label{eq:fixpnt}
	\begin{align}
	F(s_*,h_*)  & = -k_\mathrm{cat}(s_*,h_*) \, s_* + k_\mathrm{S}\, (1-s_*) = 0\,,  \label{eq:fixpnt-F} \\
	G(s_*,h_*)  & = -k \,p(s_*,h_*) \,h_* + k_\mathrm{H}\,(1-h_*) = 0\,, \label{eq:fixpnt-G}
	\end{align}
\end{subequations}
or, equivalently, by $s_* := s_\mathrm{S}^\mathrm{nc}(h_*)=s_\mathrm{H}^\mathrm{nc}(h_*)$, with the functions given in \cref{eq:nullcline-s,eq:nullcline-h}. Generally, these equations admit more than a single solution. 
The nature of each obtained fixed point then follows from the corresponding linear stability problem.

Introducing the deviation $\bm{\epsilon}=(s-s_*, h-h_*)$ from the fixed point $(s_*, h_*)$ and linearizing the flow field, \cref{eq:rre2}, around this point, one obtains the linear flow equation $d\bm{\epsilon}/dt= J(s_*,h_*) \bm{\epsilon} + O(\bm{\epsilon}^2)$ with the Jacobian $J=\partial(F,G)/\partial(s,h)$, which is a $2\times2$ matrix with elements $J_{11}=\partial F/\partial s$, $J_{12}=\partial F/\partial h$, $J_{21}=\partial G/\partial s$, $J_{22}=\partial G/\partial h$, evaluated for the solution at the point $(s_*, h_*)$. This linear ordinary differential equation is solved with the ansatz $\bm{\epsilon} \propto \exp(\lambda t)$, which leads to the characteristic polynomial $\lambda^2 - (\Tr J) \, \lambda + \Det J = 0$ in the growth rate $\lambda$.
The two roots $\lambda_\pm$ are complex-valued if the discriminant $\Delta = (\Tr J)^2 - 4 \Det J$ is negative, and real-valued otherwise. Recalling that $\Det J = J_{11} J_{22} - J_{12} J_{21} = \lambda_+ \lambda_-$ and $\Tr J = J_{11} + J_{22} = \lambda_+ + \lambda_-$, it follows that the flow near the fixed point under consideration is a saddle if $\Det J < 0$. If $\Det J > 0$, it is either a node ($\Delta > 0$), a spiral (or focus) ($\Delta < 0$), or a center ($\Delta = 0$), as the border case. The nodes and spirals can be stable ($\Tr J < 0$) or unstable ($\Tr J > 0$).
The conditions
\begin{equation}
  \Tr J =0, \quad \Det J =0, \quad \text{and} \quad \Delta =0,
  \label{eq:stability-boundaries}
\end{equation}
allow us to divide the parameter space into regions of sustained oscillations, stable steady states, and bistability.

Many of the reaction parameters are system specific and therefore remain fixed for the urea--urease reaction. What can be potentially changed in the experiment are the external concentrations $[\ce{S}_\textrm{ext}]$ and $[\ce{H+}_\textrm{ext}]$ and the reaction speed of the catalytic step, $v_\textrm{max} \propto [\ce{E}]$, which depends on the total concentration $[\ce{E}]$ of urease in the vesicle.
Furthermore, the rates $k_\textrm{H}$ and $k_\textrm{S}$ are directly related to the experimentally relevant permeabilities of the membrane to the hydrogen ion \ce{H+} and urea \ce{S} \cite{Miele:Proc2016, Miele:LNBE2018}, and it was found in a theoretical study that these rates must satisfy certain conditions for the existence of oscillations \cite{Bansagi:JPCB2014}.
Here, we consider the ratio $\bar K_\textrm{H} = k_\textrm{H}/k_\textrm{S}$, which specifies the degree of differential transport through the membrane.
In the following, we will present the overall classification of the type of behavior on the $\kappa$-$\bar K_\textrm{H}$ plane
(recall that $\kappa=[\ce{S}_\textrm{ext}]/[\ce{H+}_\textrm{ext}]$) and then show how the domain of oscillations varies upon changing the parameters $\bar v_\textrm{max}=v_\textrm{max}/(k_\textrm{S} [\ce{S}_\textrm{ext}])$ and $\bar K_\textrm{M} = K_\textrm{M}/[\ce{S}_\textrm{ext}]$.  Thereby, our analysis effectively covers the dependencies on $k_\textrm{H}$ and $k_\textrm{S}$ as well as on $[\ce{S}_\textrm{ext}]$, $[\ce{H+}_\textrm{ext}]$, and  $[\ce{E}]$.

\begin{figure}
	\centering
	\includegraphics[width=0.98\textwidth]{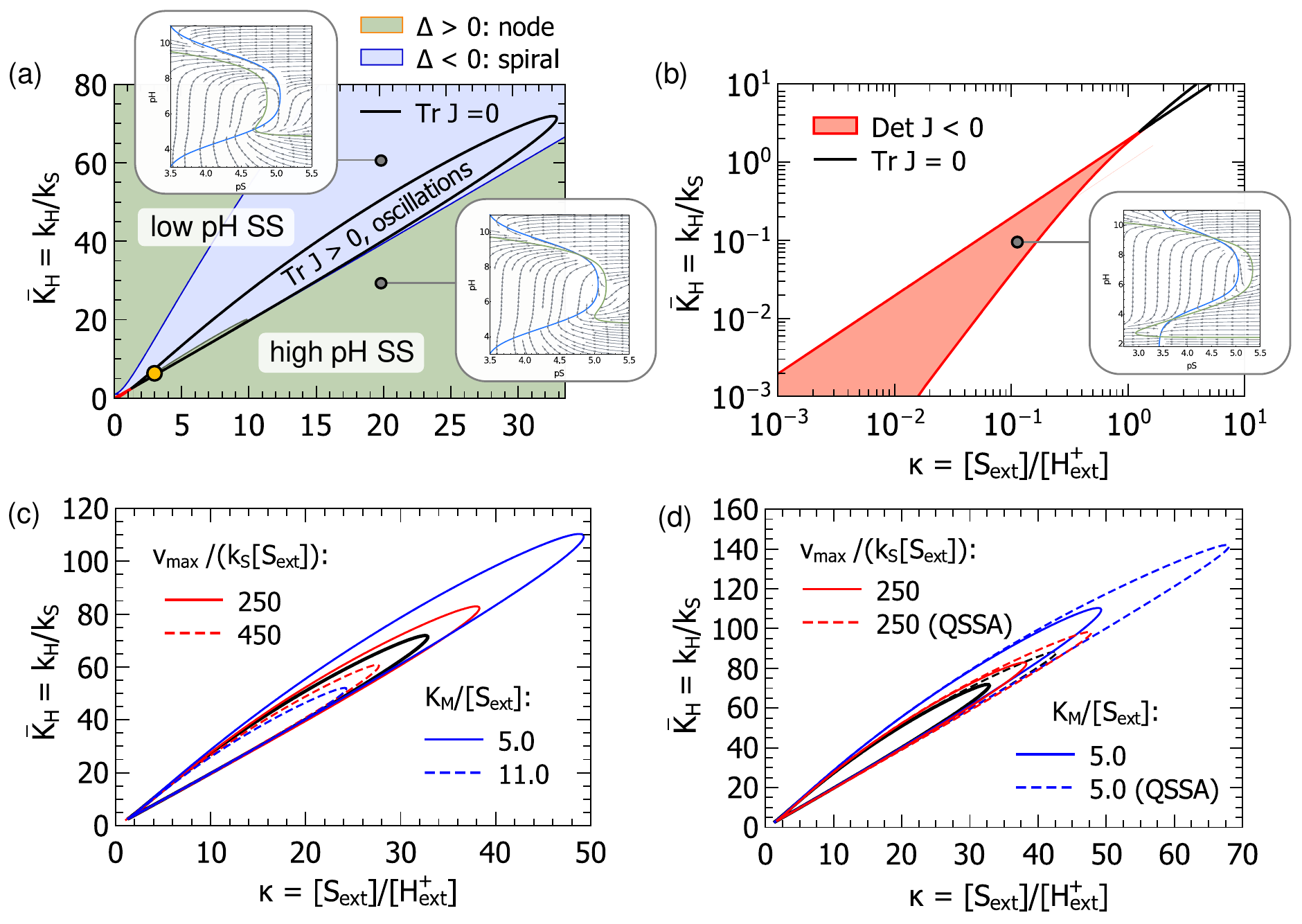}
	\caption{Stability diagrams in the $\kappa$-$\bar K_\textrm{H}$ plane plane, shown for $\bar v_\textrm{max}=347$ and $\bar K_\textrm{M} = 7.9$ if not stated differently.
	~(a)~Domains with a single fixed point of different behavior, leading to self-sustained oscillations (region within the thick black line, $\Tr J > 0$), or steady states (SS) at low or high \ce{pH} values (region outside of the thick black line, $\Tr J < 0$). The latter are approached via damped oscillations (spiral, shaded blue region, $\Delta < 0$) or without oscillations, in a fully overdamped fashion (node, shaded green region, $\Delta > 0$).
	The domains are delimited by the conditions $\Tr J = 0$ (thick black line) and $\Delta = 0$ (center, thin violet line). The orange disk indicates the set of parameters used for all other results in this work, $\kappa= 2.92$, $\bar K_\textrm{H} = 6.43$.
	The insets show phase plots and the location of the nullclines at two exemplary high and low \ce{pH} SS, at $(\kappa,K_\textrm{H})=(20,30)$ and $(20,60)$.
	~(b)~Close-up of the bottom-left corner of panel (a), revealing the domain of bistability ($\Det J < 0$) with two attractive fixed points and a saddle in between, as illustrated in the inset corresponding to $(\kappa, \bar K_\textrm{H})=(0.1,0.1)$.
	%In panels (a) and (b), the insets show the phase flow and the \ce{S} and \ce{H+} nullclines for the indicated points.
	~(c)~Domain of oscillation for different values of $v_\textrm{max}$ and $K_\textrm{M}$. The thick solid black line is the same as in panel (a) and serves as reference.
	~(d)~Comparison of the predictions of the accurate model, \cref{eq:rre2}, (solid lines) with those from invoking the QSSA for the product, \cref{eq:rre2A} (dashed lines). The thick solid black line is the reference line as in panel (a) and the dashed black line corresponds to its QSSA counterpart.
	}
	\label{fig:stability}
\end{figure}

Calculating the elements of the Jacobian $J(s_*, h_*)$ from \cref{eq:rre2}, we find after some algebra:
\begin{subequations} \label{eq:Jac}
    \begin{align}
        J_{11} % = \frac{\partial F}{\partial s}
          & = -k_\textrm{S} \frac{\bar K_\textrm{M}}{\bar K_\textrm{M} + s_*} \frac{1-s_*}{s_*} - k_\textrm{S}\,, \label{eq:Jac-J11} \\
        J_{12} % = \frac{\partial F}{\partial h}
          & = k_\textrm{S} (1 - s_*) f_\textrm{H}(h_*) z(h_*)\,, \label{eq:Jac-J12} \\
        J_{21} % = \frac{\partial G}{\partial s}
          & = -\frac{2 k_\textrm{S} \alpha_\textrm{S} h_*}{\sqrt{b_*^2 + 4c_*}} \frac{\bar K_\textrm{M}}{\bar K_\textrm{M} + s_*} \frac{1-s_*}{s_*}\,, \label{eq:Jac-J21} \\
        J_{22} % = \frac{\partial G}{\partial h}
          & = \frac{\alpha_\textrm{S} h_*}{\sqrt{b_*^2 + 4c_*}} \left[ k_\textrm{H}\frac{1-h_*}{\alpha_\textrm{S} h_*} b'_* + 2 J_{12} \right] -\frac{k_\textrm{H}}{h_*}\,, \label{eq:Jac-J22}
    \end{align}
\end{subequations} 
where $b_*=b(h_*)$, $c_*=c(s_*,h_*)$ and $b'_* = db(h_*)/dh$ are evaluated at the fixed point $(s_*,h_*)$ and we have abbreviated $z(h_*) := \bar K_\textrm{E1}^{-1}-\bar K_\textrm{E2}/h_*^2$ with $\bar K_\textrm{E1}=K_\textrm{E1} / [\ce{H+}_\textrm{ext}]$, $\bar K_\textrm{E2}=K_\textrm{E2} / [\ce{H+}_\textrm{ext}]$.
These expressions together with the conditions in \cref{eq:stability-boundaries} and the fixed point, \cref{eq:fixpnt}, define the boundaries of the various types of fixed point behavior in the $\kappa$-$\bar K_\textrm{H}$ plane (\cref{fig:stability}a).
Setting $\bar v_\textrm{max}=347$ and $\bar K_\textrm{M} = 7.90$ and selecting each of the conditions in \cref{eq:stability-boundaries}, we have numerically determined the values of $\bar K_\textrm{H}$, $s_*$, and $h_*$ that satisfy the respective condition and \cref{eq:fixpnt} for a range of prescribed values of $\kappa$.
We find that in almost the whole plane, except for a small region at low values of $\kappa$ and $\bar K_\textrm{H}$ (lower left corner of the plot), there is a single fixed point with $\Det J > 0$, meaning that the fixed point is either a node (green shaded, $\Delta > 0$) or a spiral (blue shaded, $\Delta < 0$), separated by the border case of a center (thin blue line, $\Delta = 0$). Nearly the whole domain of oscillations ($\Tr J > 0$, enclosed by the thick black line) is characterized by an unstable spiral as fixed point. Merely near the lower-$\bar K_\textrm{H}$ border for $1.5 \lesssim \kappa \lesssim 10$, there is a narrow stripe of unstable nodes. It indicates that, although less frequently observed than the spiral along with the spiral, an unstable node is also a possible type of
fixed point that can be enclosed by a limit cycle, in accordance with the Poincar{\'e}--Hopf theorem.

The regions above (higher $\bar K_\textrm{H}$) and below (lower $\bar K_\textrm{H}$) of the domain of oscillations exhibit steady states with low and high pH levels, respectively.
Here, $\Tr J < 0$ and the corresponding steady state is approached either via damped oscillations (blue shading) or aperiodically in an overdamped fashion (green shading); the domain of high pH steady states is, however, predominantly overdamped. The corresponding fixed points in the \ce{pS}-\ce{pH} plane (see insets of \cref{fig:stability}a) are located either below (low \ce{pH} steady state) or above (high \ce{pH} steady state) the neutral $\ce{pH}=7$ level; the latter is set by the apex point of the \ce{S} nullcline. Therefore, they can be referred to as the acid ($\ce{pH} < 7$) and base ($\ce{pH} > 7$) steady states, respectively.

A domain of bistability exists at small values $\kappa \lesssim 1$ and $\bar K_\textrm{H} \lesssim 1$, which is better resolved on logarithmic scales (\cref{fig:stability}b, red shading).
In this domain, the fixed point discussed so far is a saddle ($\Det J  < 0$), and the domain is delimited by the condition $\Det J  = 0$ (dark red lines); it touches the domain of oscillations ($\Tr J =0$, solid black line) at the largest values of $\kappa$ and $\bar K_\textrm{H}$ with $\Det J  = 0$ (top-right corner of \cref{fig:stability}b).
The saddle point is accompanied by two stable nodes, acting as attractors, which is exemplarily shown by the phase plot for the point $(\kappa, \bar K_\textrm{H})=(0.1,0.1)$ in the inset of \cref{fig:stability}b. The \ce{S} and \ce{H+} nullclines intersect in three different fixed points, the outermost fixed points define the low (acid) or high (base) \ce{pH} steady states. Thus, which of the two states the system evolves to depends solely on the initial state. This completes the overall structure of the possible equilibria of the urea--urease reaction. Being interested in oscillatory behavior, we will now discuss how the detected domain of oscillations transforms upon changing other control parameters.

\Cref{fig:stability}c illustrates what happens to the domain of oscillations upon changing the parameters $\bar v_\textrm{max}$ and $\bar K_\textrm{M}$. The curve shown by thick black line reproduces the domain of oscillations from \cref{fig:stability}a, for $v_\textrm{max}=347$ and $\bar K_\textrm{M} = 7.90$, and serves as a reference. First, changing only the reaction speed $\bar v_\textrm{max}$ (for $\bar K_\textrm{M} = 7.90$ fixed) leads to an elongation of the domain for smaller ($\bar v_\textrm{max}=250$, solid red line) and a a shrinking for larger ($\bar v_\textrm{max}=250$, dashed red line) values of $\bar v_\textrm{max} \propto [E]$; under these changes, the domain keeps its elongated, petal-like shape oriented approximately along the line $\bar K_\textrm{H} \approx 2\kappa$. Second, we stick to the value $v_\textrm{max}=347$ and obtain a similar effect upon changing $\bar K_\textrm{M}$: the domain of oscillations elongates at lower ($K_\textrm{M}=5$, solid blue line) and shrinks at higher ($K_\textrm{M}=11$, dashed blue line) values of $\bar K_\textrm{M}$. In all these situations, the elongation and shrinking affects the rounded end of the petal (maximal values of $\kappa$ and $\bar K_\textrm{H}$ for oscillations), while it remains unchanged at its cusp end ($\kappa \approx \bar K_\textrm{H} \approx 1$).

We now also show that the accuracy of the model plays an important role for correctly obtaining the boundaries of the oscillation domain. We compare the predictions of the accurate model \eqref{eq:rre2} with those based on the QSSA for the product, \cref{eq:rre2A}. Note that for model \cref{eq:rre2A}, the equation for $h$ is different and the elements $J_{21}$ and $J_{22}$ of the Jacobian differ. Evaluating the derivatives of the r.h.s. of \cref{eq:rre2A-h+}, we obtain
\begin{subequations} \label{eq:Jac-pqssa}
    \begin{align}
        J_{21} & = -\frac{2 k_\textrm{S} \kappa }{u(h_*)} \frac{\bar K_\textrm{M}}{\bar K_\textrm{M} + s_*} \frac{1-s_*}{s_*}\,, \label{eq:Jac-pqssa-J21} \\
        J_{22} & = \frac{1}{u(h_*)}\left[ k_\textrm{H}\frac{1-h_*}{h_* } + 2\kappa  J_{12} \right] -\frac{k_\textrm{H}}{h_*}\,. \label{eq:Jac-pqssa-J22}
    \end{align}
\end{subequations}
Alternatively, these expressions follow from \cref{eq:Jac-J21,eq:Jac-J22} by passing to the limit of large and positive $b(h_*)$, as discussed for \cref{eq:p-b-large-approx,eq:p-qssa-approx},
so that
$\sqrt{b_*^2 +4c_*} \simeq b(h_*) \simeq \alpha_\textrm{H} h_* u(h_*)$ and $db(h_*)/dh \simeq \alpha_\textrm{H}$.
In \cref{fig:stability}d, we compare the predictions for the domain of oscillations resulting from the accurate and the QSSA-based models for a few values of $v_\textrm{max}$ and $\bar K_\textrm{M}$.
We find that in all cases the QSSA significantly overestimates the size of the domain of oscillations. The domain boundaries remain close for the two models if $\kappa$ is small, but they start to deviate as $\kappa$ grows. The root of this deficiency of the QSSA-based model lies in the restricted applicability of the QSSA to low $\ce{pH}$ values:
while moving from lower to higher values of $\kappa$ along the upper (high-$\bar K_\textrm{H}$) part of the boundary of the petal-shaped domain, the fixed point lies initially in the low \ce{pH} regime,
but gradually shifts towards higher \ce{pH} levels. Upon reaching the rounded end of the petal and following the lower part of the boundary, the \ce{pH} value (and thus $h_*$) reaches and then goes beyond the point where $b(h_*) = 0$ (in case of the bold black reference curve, this happens at $\ce{pH} \approx 6.17$) and the QSSA model fails.

\subsection{Analytic results for the fixed point governing the oscillations}

The fixed point $(s_*, h_*)$ is determined by the condition $s_* := s_\mathrm{S}^\mathrm{nc}(h_*)=s_\mathrm{H}^\mathrm{nc}(h_*)$, with the functions given in \cref{eq:nullcline-s,eq:nullcline-h}. To make analytic progress, we
observe that $\beta(h)^2 > (\bar K_M - 1)^2 \gg \bar K_M$ for all \ce{pH} values provided that $K_M \gg [\ce{S}_\mathrm{ext}]$, which permits one to approximate \cref{eq:nullcline-s} by $s_\mathrm{S}^\mathrm{nc}(h) \simeq K_\textrm{M}/\beta(h)$. As a result, the fixed point condition reads
\begin{align}
    \bar K_\mathrm{M} + \bar v_\mathrm{max} f_\mathrm{H}(h_*)
    \simeq \frac{2 \kappa k_\textrm{S} \bar v_\mathrm{max}f_\mathrm{H}(h_*)} {k_\mathrm{H} (1-h_*)u(h_*)} \,.  \label{eq:fixedpoint-eq}
\end{align}
For low $\ce{pH}$ level, we can further approximate $u(h) \simeq 1$ and $f_\textrm{H}(h) \simeq (1 + h/\bar K_\textrm{E1} )^{-1}$,
which holds for $[\ce{H+}] \gg 1/k'$ and $[\ce{H+}] \gg K_\textrm{E1}$, respectively.
With this, \cref{eq:fixedpoint-eq} transforms to the quadratic equation $h_*^2+B h_*- C \simeq 0$ with
coefficients $B=\bar K_\textrm{E1}(1+\bar v_\textrm{max}/\bar K_\textrm{M})-1$ and
$C=B+1-2\kappa k_\textrm{S} \bar K_\textrm{E1} \bar v_\textrm{max} /(k_\textrm{H} \bar K_\textrm{M})$;
only the positive root is physically relevant:
\begin{align}
    h_{\ast} \simeq -\frac{B}{2} + \frac{1}{2}\sqrt{B^2+4C} \,.  \label{eq:fixedpoint}
\end{align}
For the chosen set of parameters (orange disk in \cref{fig:stability}a), these approximations yield the fixed point at $(\ce{pS}_{\ast},\ce{pH}_{\ast}) \approx (4.314,4.575)$, which is very close to the earlier finding \cite{Straube-etal:JPCL2021} $(\ce{pS}_{\ast},\ce{pH}_{\ast}) \approx (4.311, 4.569)$ from the numerical solution of the full equation $s_\mathrm{S}^\mathrm{nc}(h_*)=s_\mathrm{H}^\mathrm{nc}(h_*)$.

From the Poincar\'{e}-Hopf theorem, we have concluded earlier (see discussion of \cref{fig:phase-plot}b) that this fixed point is either a node or a spiral. More explicitly, we calculate the Jacobian $J$ at this point. The elements $J_{21}$ and $J_{22}$ can be obtained within the QSSA, \cref{eq:Jac-pqssa-J21,eq:Jac-pqssa-J22} and additionally $u(h_*)=1$, since the fixed point is in the acid regime (in particular, $\ce{pH} < 6.17$ so that $b(h_*) > 0$); the elements $J_{11}$ and $J_{12}$ are given by \cref{eq:Jac-J11,eq:Jac-J12}. We find $\Tr J \approx 9.6 \times 10^{-3} >0$, $\Det J \approx 5.6\times 10^{-5} >0$ and $\Delta \approx -1.9\times 10^{-4} <0$, and these values confirm that the fixed point is of the spiral type.

\section{Influence of water ionization}

We now investigate the role of water ionization accounted for previously in the literature\cite{Bansagi:JPCB2014}. We will show that
although the exact mapping from the three- to the two-variable models admits a straightforward generalization, it brings no essential advantages over its simpler counterpart, \cref{eq:rre2}. Concerning the less accurate reduction based on the QSSA for the product \ce{P}, the inclusion of water ionization removes the degeneracy exhibited by model \eqref{eq:rre2A}. However, the qualitative structure of the flow field, the nature of the limit cycle at high \ce{pH} and the resulting oscillations deviate considerably from those of the exact reduction scheme.

To account for water ionization, we supplement the in-volume reactions \eqref{4sm} by the auto-dissociation reaction
\begin{equation}
  \ce{H+ + OH- <=> H2O} \,.
\end{equation}
Then, the corresponding reaction reaction rate equations involve five variables, see e.g. Eqs. (A3)--(A7) in Ref.~\citenum{Bansagi:JPCB2014}. The scheme further assumes the exchange of \ce{OH-} with the reservoir at, for simplicity, the same rate $k_\textrm{H}$ as for the \ce{H+} exchange. Assuming instantaneous equilibrium between \ce{H2O}, \ce{H+} and \ce{OH-}, 
we exclude the latter species as a slaved one by means of the relation $[\ce{OH-}](t)=K_\textrm{W}/[\ce{H+}](t)$ with $K_\mathrm{W}=\SI{e-14}{M}$. Further, by applying \cref{eq:qssa-ph+} to eliminate $\ce{PH+}$ and proceeding to dimensionless variables, we arrive at the three-variable generalization of \cref{eq:rre3nondim}:
\begin{subequations} \label{eq:rre3A}
	\begin{align}
	\frac{ds}{dt} & = -k_\mathrm{cat}(s,h) s + k_\mathrm{S} (1 - s)\,,  \label{eq:rre3A-s} \\
%	\frac{dh}{dt} & = \left[-k \,p \,h + k_\mathrm{H}(1 - h) w_1(h)\right] w_2^{-1}(h)\,, \label{eq:rre3A-h+} \\
	w_2(h)\frac{dh}{dt} & = -k \,p \,h + k_\mathrm{H}(1 - h) w_1(h)\,, \label{eq:rre3A-h+} \\
	\frac{dp}{dt} & = 2 \alpha_{\textrm{S}} k_\mathrm{cat}(s,h) s - \alpha_{\textrm{H}} k p h  u(h)\,.  \label{eq:rre3A-p}
	\end{align}
\end{subequations}
Here, the auxiliary functions $w_1(h) = 1 + K_\mathrm{W} / ([\ce{H+}_\mathrm{ext}]^2 h)$ and $w_2(h) = 1 + K_\mathrm{W} / ([\ce{H+}_\mathrm{ext}]^2 h^2)$ account for the water ionization. They are removed from the equations upon letting $K_\mathrm{W} \to 0$, which yields $w_1(h)=w_2(h)=1$, so that model \eqref{eq:rre3A} reduces to the simpler case, \cref{eq:rre3nondim}.

This generalized model admits an exact reduction to two variables subject to the constraint \eqref{eq:elimcond} and after the elimination of $p$. The derivation repeats all the steps above for the case without water ionization, yielding
\begin{subequations} \label{eq:rre2-gen}
	\begin{align}
	\frac{ds}{dt} & = -k_\mathrm{cat}(s,h) s + k_\mathrm{S} (1 - s)\,,  \label{eq:rre2gen-s} \\
%	\frac{dh}{dt} & = \left[-k p(s,h) h + k_\mathrm{H}(1 - h) w_1(h)\right] w_2^{-1}(h)\,. \label{eq:rre2gen-h+} 
 	w_2(h)\frac{dh}{dt} & = -k p(s,h) h + k_\mathrm{H}(1 - h) w_1(h)\,. \label{eq:rre2gen-h+}
	\end{align}
\end{subequations}
Here, $p=p(s,h)$ assumes the same form as before, \cref{eq:p}, with the generalized coefficients
%\cref{eq:cffs-b,eq:cffs-c}
\begin{subequations} \label{eq:cffs-bc-gen}
	\begin{align}
	b(h)    & =  \alpha_\textrm{H} h u(h) w_2(h) + (1 - h^{-1})k_\mathrm{H}/k\,, \label{eq:cffs-bc-gen-b} \\
	c(s,h)  & = \frac{2 \alpha_\textrm{S} k_\mathrm{cat}(s,h) w_2(h)}{k} \, s \: \ge 0\,. \label{eq:cffs-bc-gen-c}
	\end{align}
\end{subequations}

Instead of eliminating the variable $p$ exactly, we can impose the QSSA for $p$ as above for the case without water ionization. Note that the rate equation for \ce{P} preserves its form irrespective of whether water ionization is accounted for (\cref{eq:rre3A-p}) or disregarded (\cref{eq:rre3nondim-p}). For this reason, \cref{eq:kph} is recovered upon setting $dp/dt=0$. Inserting it into \cref{eq:rre2gen-h+}, we find the generalization of model \eqref{eq:rre2A}, now including the effects of water ionization:
\begin{subequations} \label{eq:rre2B}
	\begin{align}
	\frac{ds}{dt} & = -k_\mathrm{cat}(s,h) s + k_\mathrm{S} (1-s)\,,  \label{eq:rre2B-s} \\
%	\frac{d h}{dt} & = \left[-2 \kappa k_\mathrm{cat}(s,h) s\, u^{-1}(h) + k_\mathrm{H}(1-h)w_1(h)\right] w_2^{-1}(h)\,. \label{eq:rre2B-h+} 
 	w_2(h)\frac{d h}{dt} & = -\frac{2 \kappa k_\mathrm{cat}(s,h)}{u(h)} s  + k_\mathrm{H}(1-h)w_1(h) \,. \label{eq:rre2B-h+}
	\end{align}
\end{subequations}
This is essentially the model originally suggested by \citet{Bansagi:JPCB2014}, who obtained it in the limit $\alpha_\textrm{H} h \gg 1$ corresponding to $u(h) = 1$, which is a very reasonable approximation for $\ce{pH} \lesssim 9$ given that $\alpha_H \approx 2\times 10^{6}$. Note that the equations for the substrate \ce{S} are identical to  \cref{eq:rre2-s}, and hence the \ce{S} nullcline is the same for all two-variable models and is given by \cref{eq:nullcline-s}. In full analogy with the case without water ionization, the \ce{H+} nullclines of models \eqref{eq:rre2-gen} and \eqref{eq:rre2B} are the same and are given by
\begin{align}
s_\mathrm{H}^\mathrm{nc}(h)=
\bar K_\mathrm{M} \left[\frac{2 \kappa k_\mathrm{S} \bar v_\mathrm{max}f_\mathrm{H}(h)}
{k_\mathrm{H}(1-h) w_1(h) u(h)}-1\right]^{-1}\,.
\label{eq:nullcline-h-B}
\end{align}

The impact of water ionization on the phase flow and the limit cycle is shown in \cref{fig:phase-plot}.
First, we find that the accurate reductions \eqref{eq:rre2} and \eqref{eq:rre2-gen} of the three-variable models give very close results for both the flow field and the limit cycle (see \cref{fig:phase-plot}a): there is merely a tiny shift of the \ce{H+} nullclines of the two models, cf. green and dashed dark blue lines given by \cref{eq:nullcline-h,eq:nullcline-h-B}, respectively. This indicates that the effects of water ionization on the oscillation dynamics of the full model are negligible.
However, for the two QSSA-based reductions, models \eqref{eq:rre2A} and \eqref{eq:rre2B}, the obtained phase flows are quite different (\cref{fig:phase-plot}c and d), despite the nullclines being almost identical.
In particular, including the water ionization leads to a regularization of the flow in the vicinity of the high-\ce{pH} branch of the \ce{H+} nullcline; as a consequence, the high-\ce{pH} part of the limit cycle is more rounded.
Moreover, the flow field in panel~(d) appears similar to the one of the exact model (panel~(a)),
in particular, in comparison to panel~(c).
On the other hand, despite the regularization the shape of the limit cycle of model \eqref{eq:rre2B} is still dictated by the \ce{H+} nullcline (panel~(d)) rather than the \ce{S} nullcline (panel~(a)), as required by the exact models. As for the QSSA-based model \eqref{eq:rre2A}, this inconsistency leads similarly to an underestimate of the oscillation period by nearly $30\%$, namely \SI{12.7}{\minute} to be compared to the reliable prediction of \SI{17.6}{\minute} within model \eqref{eq:rre2}, and an overestimate of the amplitude of \ce{pH} oscillations, $\max(\ce{pH})- \min(\ce{pH}) \approx 4.8$ vs.\ $4.0$ from the accurate model.
Thus, explicitly accounting for the water ionization in the model mitigates the issues arising from the QSSA for the product only apparently.

\section{Discussion}

We have theoretically studied an urea-urease-based pH oscillator confined to a giant lipid vesicle, which is capable of differential transport of urea and hydrogen ion across the uni\-lamellar membrane and serves as an open reactor. In contrast to conventional pH oscillators in closed chambers, the exchange with the vesicle's exterior periodically resets the pH clock that switches the system from acid to basic. Here, we have focused on large vesicles of sizes of several micrometers, which justifies a deterministic treatment of the dynamics. Quite importantly, as shown recently by a stochastic simulation study \cite{Straube-etal:JPCL2021}, the structure of the limit cycle of the deterministic rate equations controls not only the behavior for giant vesicles, but also dominates the pronouncedly stochastic oscillations in vesicles of submicrometer size. This has prompted the detailed analysis of the structure of the phase flow and the limit cycle.

Starting from a reaction scheme involving four species, namely urea as the substrate \ce{S}, hydrogen ion \ce{H+}, ammonia as product \ce{P}, and ammonium as its ion form \ce{PH+}, we have obtained accurate reduced models by eliminating the two product species. We have first reduced the system of rate equations to one in three variables (\cref{eq:rre3}) by imposing a QSSA on the dynamics of \ce{PH+} (\cref{eq:qssa-ph+}) and thus removing it from the equations, which is justified by a timescale separation. Next, we have exploited the scale separation in the oscillation amplitude of \ce{PH+} compared to \ce{P} and \ce{H+}, which allows one to eliminate \ce{P} and to arrive at model \eqref{eq:rre2} for two variables, \ce{S} and \ce{H+}, which is a virtually exact representation of the three-variable model. In particular, the latter step introduces a tight constraint that couples the dynamics of \ce{H+} and \ce{P}, implying that the three-variable model is effectively a two-dimensional one. The constraint manifests itself as a non-trivial manifold on which the phase flow and the limit cycle are restricted to exist (\cref{fig:phaseplot3d}). The structure of the phase flow and the properties of the limit cycle are best uncovered using a logarithmic representation---in contrast to conventional examples of oscillators.

By analyzing the phase flow and the mutual positioning of the nullclines, we have shown that the limit cycle is governed by the \ce{S} nullcline. Noteworthy, this outcome is in contrast to our expectations based on the QSSA for \ce{P} from a previous study, which suggested that the limit cycle is set by the \ce{H+} nullcline \cite{Straube-etal:JPCL2021}.
We have further demonstrated that the QSSA for \ce{P} approximates the full model only in the acid regime (low \ce{pH}) and leads to degenerate behavior of the flow field near the high-\ce{pH} branch of the \ce{H+} nullcline. Even though the nullclines remain identical for both two-dimensional models, the structure of the phase flow is different, in particular, in the basic regime. Moreover, the QSSA for \ce{P} is mathematically unjustified and leads to inconsistencies in the regime of high \ce{pH}. Thus, the quality of the model and the accuracy of its predictions, including the oscillation period, are highly sensitive to the choice of the reduction scheme.

For the experimentally relevant set of parameters studied here, the dynamics has a single fixed point as exhibited by the single intersection of the nullclines. According to the Poincar{\'e} index theory or, more generally, the Poincar{\'e}--Hopf theorem, it follows that the limiting periodic trajectory must enclose this fixed point, which must be either a node, a spiral, or a center.
To gain insights into the parameter region where sustained oscillations occur, we have performed a linear stability analysis around the fixed point and determined the domain of oscillations in the $\kappa$-$\bar K_\textrm{H}$ plane in terms of the
dimensionless ratios $\kappa=[\ce{S}_\textrm{ext}] / [\ce{H+}_\textrm{ext}]$ and $\bar K_\textrm{H}= k_\textrm{H}/k_\textrm{S}$.
The ratio $\kappa$ is controlled in the experiment by defining the external concentrations, and $\bar K_\textrm{H}$ is a direct measure of the differential permeability of the membrane to the hydrogen ion and the substrate molecule urea. Oscillations can exist only if both ratios fall within a certain range of values; outside of this parameter domain, the system evolves to either an acid (low \ce{pH}) or base (high \ce{pH}) steady state.

The domain of oscillations was found to have an elongated petal-like shape, which is narrow for small values of $\bar K_\textrm{H}$ and widens for larger values (\cref{fig:stability}). Our results emphasize the importance of the differential transport for the \ce{pH} oscillations and a precise control of the external concentrations. More robust oscillations may be observed experimentally if the parameter $\bar K_\textrm{H}$ can be increased, which may be achieved by changing the temperature (assuming an Arrhenius behavior of the permeabilities) or by enclosing the reaction in bi-lamellar vesicles (which would approximately turn $\bar K_\textrm{H}$ for a single bilayer into $\bar K_\textrm{H}^2$). Changing other system parameters, such as the urease concentration in the vesicle (which is proportional to the speed of the catalytic step), leads to an elongation or an shrinking of the oscillation domain, essentially without changing its shape. As a caveat, we note that relying on the QSSA for the product generally overestimates the tendency of the reaction system to oscillate and yields a too large domain of oscillations.

We have further investigated the concequences of explicitly accounting for water ionization in the models, in particular, in combination with the QSSA-based reduction for product. First, we found that this extension of the full model has a negigible effect on the oscillation dynamics, neither qualitatively nor quantitatively. Second, we concluded that the reduction scheme based on the QSSA for the product leads to a model with deficiencies also when water ionization is included. On one hand, it removes the degenerate flow behavior in the vicinity of the \ce{H+} nullcline for high \ce{pH} values. On the other hand, it still predicts the \ce{H+} nullcline to act as an attractor that sets the limit cycle, which is qualitatively wrong (\cref{fig:phase-plot}d). As a result, the quantitative predictions of the period and amplitude of the \ce{pH} oscillations remain unsatisfactory despite the regularization. Both QSSA-based schemes for the product, with and without water ionization, underestimate the period of oscillation by nearly $30\%$ and overestimate the amplitude by around $20\%$ compared to the accurate model.

Eventually, whereas the two-variable model is more amenable to analytic treatments, its three-variable counterpart admits for a meaningful interpretation as a reaction scheme and is favorable for stochastic simulations. These models can be used for accurate descriptions of \ce{pH} oscillations in giant, but also small vesicles; for the latter, correctly reproducing the oscillation period is vital for the sound interpretation of experiments. Given the narrow parameter domain where stable oscillations exist, reliable predictions of the reaction kinetics are indispensible for the design of experiments that demonstrate the oscillatory behavior. Furthermore, a faithful model of a single \ce{pH} oscillator is a crucial prerequisite for understanding communication of vesicles and synchronization of rhythms \cite{Pikovsky-etal:2001book, Safonov-Vanag:PCCP2018, Budroni-etal:JPCL2020, Budroni-etal-PCCP2021, Miele-etal:JPCL2022}.

\begin{acknowledgement}
We thank Michael Zaks for helpful discussions. 
This research has been supported by Deutsche Forschungsgemeinschaft (DFG) through grant SFB~1114, project no.\ 235221301 (sub-project C03) and under Germany's Excellence Strategy -- MATH+ : The Berlin Mathematics Research Center (EXC-2046/1) -- project no.\ 390685689 (subprojects AA1-1 and AA1-18).
\end{acknowledgement}

\bibliography{references}

\end{document}